\documentclass[journal]{IEEEtran}
\usepackage[colorlinks,linkcolor=cyan,anchorcolor=cyan,citecolor=cyan,bookmarks=true]{hyperref}  
\usepackage[T1]{fontenc}
\ifCLASSINFOpdf
\usepackage[pdftex]{graphicx} 
\DeclareGraphicsExtensions{.eps,.pdf,.jpeg,.png}
\else
\usepackage[dvips]{graphicx}
\fi
\usepackage[compress,nospace]{cite}
\usepackage[cmex10]{amsmath}
\usepackage{epstopdf,amsthm,stfloats,siunitx,amssymb,wasysym,algorithm,algorithmic,array,url, color,subfigure} 
\usepackage{footnote}
\makesavenoteenv{tabular}
\newtheorem{theorem}{Proposition}

\usepackage{soul}

\begin{document}

%% Title and Authors
\title{Efficient DOA Estimation Method for Reconfigurable Intelligent Surfaces Aided UAV Swarm}

\author{Peng~Chen,~\IEEEmembership{Member,~IEEE},
	Zhimin~Chen,~\IEEEmembership{Member,~IEEE},
	Beixiong~Zheng,~\IEEEmembership{Member,~IEEE},
	% Zhenxin~Cao,~\IEEEmembership{Member,~IEEE}, 
	%Xiaoye Wu, 
	Xianbin~Wang,~\IEEEmembership{Fellow,~IEEE}
	%Yi Jin
	%, Can Zhu
	%\thanks{C.~Can is with NO.724 Research institute of CSIC, Nanjing 211153, China (email: ZCan1021@163.com).}
	\thanks{This work was supported in part by the National Natural Science Foundation of China (Grant No. 61801112), the Equipment Pre-Research Field Foundation, the Industry-University-Research Cooperation Foundation of The Eighth Research Institute of China Aerospace Science and Technology Corporation (Grant No. SAST2021-039), the National Key R\&D Program of China (Grant No. 2019YFE0120700), and the Natural Science Foundation of Jiangsu Province (Grant No. BK20180357). \textit{(Corresponding author: Zhimin Chen)}}
	\thanks{P.~Chen is with the State Key Laboratory of Millimeter Waves, Southeast University, Nanjing 210096, China (email: chenpengseu@seu.edu.cn).}
	%\thanks{X.~Wu is with the Beijing Aerospace TT\&C Technology Co., Ltd., China.}
	\thanks{Z.~Chen is with the School of Electronic and Information, Shanghai Dianji University, Shanghai 201306, China, and also with the Department of Electronic and Information Engineering, The Hong Kong Polytechnic University, Hong Kong (e-mail: chenzm@sdju.edu.cn). }
	\thanks{B. Zheng is with the Department of Electrical and Computer Engineering, National University of Singapore, Singapore 117583 (email: elezbe@nus.edu.sg).}
	\thanks{X.~Wang is with the Department of Electrical and Computer Engineering, Western University, Canada (e-mail: xianbin.wang@uwo.ca).}
	%\thanks{X.~Wang is with the Department of Electrical and Computer Engineering, Western University, Canada (e-mail: xianbin.wang@uwo.ca).}
	%\thanks{Y.~Jin is with Xi'an branch of China Academy of Space Technology, Xi'an 710100, China (email: john.0216@163.com).} 
}

\markboth{IEEE Transactions on Signal Processing}%
{Shell \MakeLowercase{\textit{et al.}}: Bare Demo of IEEEtran.cls for Journals}

\maketitle
\begin{abstract}
	The conventional direction of arrival (DOA) estimation methods are performed with multiple receiving channels. In this paper, a changeling DOA estimation problem is addressed in a different scenario with only one full-functional receiving channel. A new unmanned aerial vehicle (UAV) swarm system using multiple lifted reconfigurable intelligent surface (RIS) is proposed for the DOA estimation.  The UAV movement degrades the DOA estimation performance significantly, and the existing atomic norm minimization (ANM) methods cannot be used in the scenario with array perturbation.  Specifically, considering the position perturbation of UAVs, a new atomic norm-based DOA estimation method is proposed, where an atomic norm is defined with the parameter of the position perturbation. Then, a customized semi-definite programming (SDP) method is derived to solve the atomic norm-based method, where different from the traditional SDP method, an additional transforming matrix is formulated. Moreover, a gradient descent method is applied to refine the estimated DOA and the position perturbation further. Simulation results show that the proposed method achieves much better DOA estimation performance in the RIS-aided UAV swarm system with only one receiving channel than various benchmark schemes.
\end{abstract}


\section{Introduction} \label{sec1}
The direction of arrival (DOA) estimation problem has been studied for decades~\cite{9541018,9384289}. Generally speaking, existing DOA estimation methods can be roughly classified into the Fourier transformation (FT) based methods and the super-resolution methods. In general, the angular resolution of the FT-based techniques is limited by \emph{Rayleigh criterion}, and those methods with better resolution performance than the Rayleigh criterion are referred to as super-resolution methods. The most popular super-resolution methods are based on the sub-space theory, such as the multiple signal classification (MUSIC)~\cite{6747980} and the estimation of signal parameters via rotational invariant technique (ESPRIT) methods~\cite{1395953,7226785}.

Due to its exciting capability of reflecting received signals with controllable reflection amplitudes and phases, reconfigurable intelligent surface (RIS)~\cite{9174801,9309152} has been studied recently for enhancing wireless communication and DOA estimation. In~\cite{9201413}, RIS orientation and location are optimized to maximize the coverage. In~\cite{9144510}, the channel distributions of a dual-hop RIS-aided scheme and a transmitting scheme are derived. Additionally, for the RIS-assisted unmanned aerial vehicle (UAV) communication systems, the RIS is used to reflect the signals transmitted from the ground source to a UAV, for which the corresponding outage probability, average bit error rate, and average capacity are given in~\cite{9124704}.  In~\cite{9354904}, a non-iterative two-stage channel estimation framework is proposed for the DOA estimation in the point-to-point RIS-aided millimeter-wave (mmWave) multi-input and multi-output (MIMO) system. A novel direction-finding system is studied in this paper with RIS characteristics, and a corresponding high-resolution DOA estimation method will be proposed.

Recently, for the high-resolution DOA estimation techniques, sparse reconstruction-based methods have been proposed to exploit the signal's sparsity in the spatial domain and achieve better performance than the sub-space methods. Among those methods, compressed sensing (CS)-based methods~\cite{9359337,9296231} are essential to realizing the sparse reconstruction. Specifically, various CS-based methods, such as orthogonal matching pursuit  (OMP) method~\cite{4385788,1337101}, generalized belief propagation (GBP) method~\cite{1459044} and approximate message passing (AMP) method~\cite{6556987}, discretize the spatial angle into grids. However, in practice, the target cannot be on the grid exactly, which causes the \emph{off-grid} problem. To solve this problem, gridless methods have been proposed for the DOA estimation and can avoid the \emph{off-grid} problem~\cite{9016105,7313018}. The atomic norm minimization (ANM)  is a widely adopted gridless method~\cite{6576276,7307118,7091875,8432470,7917313,9146196,7314978}. In~\cite{6998075}, with mild spectral separation condition, it proves that all the frequencies can be estimated precisely by solving an ANM program in the two-dimensional scenario. In~\cite{9384289}, an irregular Toeplitz matrix and an irregular Vandermonde decomposition are given for the ANM-based DOA estimation.

For practical systems, system model errors must also be considered in spatial filtering techniques for the DOA estimation. For example, a fault detection filter design problem for a class of nonhomogeneous higher-level Markov jump systems with uncertain transition probabilities is investigated in~\cite{9447907} based on the interval type-2 fuzzy method. A robust iterative learning control (ILC) algorithm is derived based on iteratively solving a convex optimization problem formulated by a worst-case norm-optimal problem in~\cite{iet-cta}. For a nonlinear Markov jump system, ref.~\cite{rnc} converts the nonlinear terms into linear forms using the neural network linear differential inclusion techniques. Additionally, an asynchronous fault detection filter is given in~\cite{CHENG2021107353}, and sufficient conditions for the stable resultant Markov jump systems are devised.

Moreover, to achieve a better DOA estimation performance and reduce the number of receiving channels, sparse arrays have been proposed, such as the nested array~\cite{5456168}, coprime array~\cite{9411879,9367250}, etc. In~\cite{9442317}, a dilated array is presented and applies the dilation method to other array geometries on a moving platform. Furthermore, for the DOA estimation, some methods have been proposed to reduce the computational complexity. For example, in~\cite{9432742}, a neural network combining with gradient steps on the likelihood function is offered for the DOA estimation and the model order selection. Then, a  subarray sampling approach for DOA estimation is proposed, where the covariance matrix of the whole array is estimated from the subarrays using a neural network~\cite{9400719}. An recursive order method is applied for the DOA estimation with an unknown number of targets and combines two spatial modified Yule-Walker systems in~\cite{9442863}. In~\cite{9410582}, a harmonic retrieval joint multiple regression method is proposed for the DOA estimation against unknown spatially colored noise in the radar system.

In this paper, a novel direction-finding technique is proposed in a system combining UAV swarm and RIS. The implementation cost is significantly reduced due to the simplified requirement of only one full-functional receiving channel. Then, based on the system model, an ANM method is proposed for the DOA estimation to exploit the target sparsity in the spatial domain. The corresponding semi-definite programming (SDP) method is derived to solve the ANM problem. Unlike the existing techniques, the position perturbations caused by the UAVs are considered, and a transformation matrix is introduced. Moreover, a gradient descent method is also given to refine the estimated DOA and positional perturbation further. Simulations results are presented and compared with the existing approaches to show the superiority of the proposed scheme. The contributions of this paper are given as follows: 
	\begin{itemize}
		\item \textbf{A novel low-cost direction-finding system} is proposed using a UAV swarm, where the RIS elements are mounted on UAVs, and only one full-functional receiving system is used at the center UAV. Multiple measurements are realized by changing the reflected amplitudes and phases in the RIS.   
		\item \textbf{A novel DOA estimation method based on the ANM} is proposed. With the position perturbation in UAV, the existing ANM methods cannot be used directly, so we introduce the position perturbation vector and redefine a new type of atomic norm. Then, a transformation matrix is proposed for the sparse reconstruction, and ANM-based methods are extended to more general applications. 
		\item \textbf{A perturbation estimation method for the UAV movement} is proposed, which is inspired by the nonconvex optimization. Since the UAV movement degrades the DOA estimation performance, the proposed perturbation estimation method can significantly improve the DOA estimation performance. Finally, \textbf{a low bound of the DOA estimation} is also derived as a benchmark.
	\end{itemize}

The remainder of this paper is organized as follows. The novel direction-finding system is given in Section~\ref{sec2}. Then, the super-resolution DOA estimation method with position perturbation is proposed in Section~\ref{sec3}. The DOA estimation bound is derived in Section~\ref{sec4}. The simulation results are carried out in Section~\ref{sec5}. Finally,  Section~\ref{sec6} concludes  the paper.

\textit{Notations:} Upper-case and lower-case boldface letters denote matrices and column vectors, respectively. The  matrix transpose and the  Hermitian transpose are denoted as $(\cdot)^\text{T}$ and $(\cdot)^\text{H}$, respectively.  $\operatorname{diag}\{\boldsymbol{a}\}$ returns a diagonal matrix with the elements in $\boldsymbol{a}$ as its main diagonal entries. $a^*$ denotes the conjugate of $a$. The Kronecker product is denoted as $\otimes$. $\mathcal{R}\{\cdot\}$ denotes the real part of a complex value.  $\text{Tr}\{\cdot\}$ is the trace of a matrix.    $\text{vec}\{\boldsymbol{A}\}$ denotes the vectorization of $\boldsymbol{A}$ by stacking its columns into a column vector. $\|\cdot\|_2$ is  the $\ell_2$ norm.  $\odot$ denotes the element-wise product (Hadamard product).

\section{Direction Finding System Model Using UAV Swarm}\label{sec2}

\begin{figure}
	\centering
	\includegraphics[width=3in]{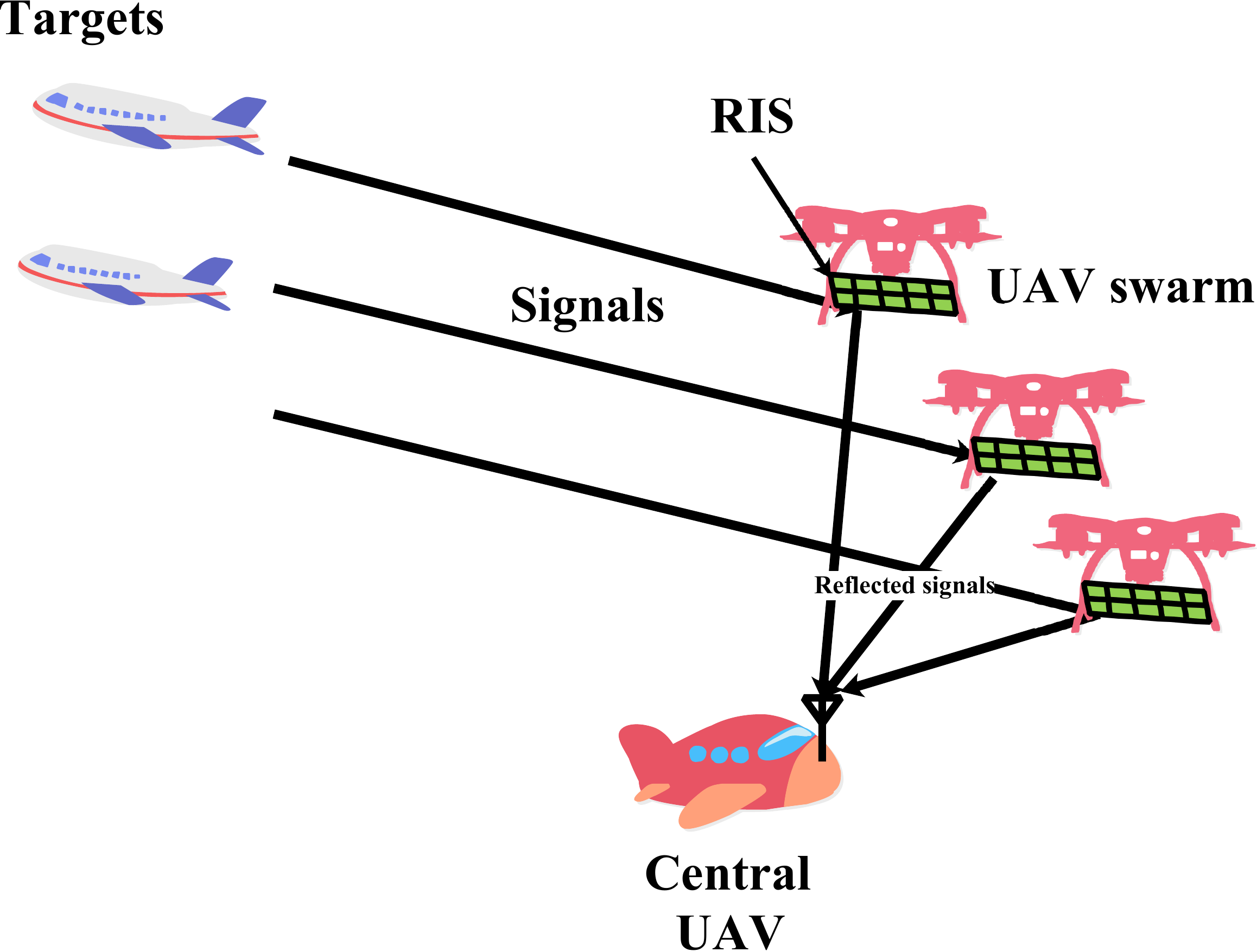}
	\caption{The direction finding system using UAV swarm and RIS.}
	\label{system}
\end{figure}

Considering a direction-finding system using $N$ UAVs, a RIS element is mounted on each UAV. Those UAVs form a uniform linear array (ULA), and a center UAV receives the reflected signals from RIS elements, as shown in Fig.~\ref{system}. There are $K$ far-field target signals, and the $k$-the signal is denoted as $s_k(t)$ with $k=0,1,\dots, K-1$.   We try to use this low-cost system to estimate the directions of the received signals, where only one full-functional receiving system is equipped at the center UAV, and  low-cost RIS elements are mounted on the considered UAV swarm.

The position of the $n$-th ($n=0,1,\dots, N-1$) UAV is denoted as $d_n$, where we take  the position of the $0$-th UAV as a reference position, i.e., $d_0=0$. Collect these positions into a vector, and we have $\boldsymbol{d}\triangleq \begin{bmatrix}
d_0,d_1,\dots,d_{N-1}
\end{bmatrix}^{\text{T}}$. Additionally, this system model can be easily extended to the two-dimensional DOA estimation. The distance between adjacent UAVs is about half of the carrier wavelength. For the $n$-th UAV, the received signal in RIS element can be expressed as
\begin{align}
x_n(t) = \sum^{K-1}_{k=0}s_k(t)e^{j2\pi\frac{d_n}{\lambda}\sin\theta_k},
\end{align}
where $\lambda$ denotes the wavelength, and $\theta_k$ is the direction of the $k$-th signal. After the RIS reflection, the received signal during the $m$-th ($m=0,1,\dots,M-1$) time slot $mT$ in the central UAV is
\begin{align}
r(mT) = \sum^{N-1}_{n=0}x_n(mT)A_{n,m}e^{j\phi_{n,m}}e^{j2\pi\frac{d_n}{\lambda}\sin\psi}+w(mT),
\end{align}
where $T$ is the period of the time slot. $A_{n,m}$ and $\phi_{n,m}$ denote the amplitude and phase caused by the $n$-th RIS element during the $m$-th time slot, respectively. $\psi$ is the known far-field angle between the central UAV and the UAV  swarm. $w(mT)\in\mathbb{C}$ denotes the additive white Gaussian noise (AWGN) with the variance being $\sigma^2_{\text{w}}$.

Only one receiving channel is used in the proposed system for the DOA estimation, which is different from the exiting system using multiple channels. Hence, the existing DOA estimation methods cannot be used directly. A DOA estimation method for the one-channel system is proposed, where the multiple measurements are received by changing the reflected signals from RIS. Collect these received signals during different time slots into a vector, and the received signal in the central UAV can be simplified as
\begin{align}
\boldsymbol{r} & \triangleq \big[r(0), r(T),\dots,r[(M-1)T]\big]^{\text{T}}\in\mathbb{C}^{M\times 1}
\notag                                                                                                                                                        \\
& = \sum^{N-1}_{n=0} \sum^{K-1}_{k=0} e^{j2\pi\frac{d_n}{\lambda}(\sin\theta_k+\sin\psi)} \begin{bmatrix}
s_k(0)
A_{n,0}e^{j\phi_{n,0}} \\
\vdots                 \\
s_k(mT)
A_{n,m}e^{j\phi_{n,m}} \\
\vdots
\end{bmatrix}+\boldsymbol{w}
\notag                                                                                                                                                        \\
& \approx \sum^{N-1}_{n=0}  \begin{bmatrix}
A_{n,0}e^{j\phi_{n,0}} \\
\vdots                 \\
A_{n,m}e^{j\phi_{n,m}} \\
\vdots
\end{bmatrix} \sum^{K-1}_{k=0} e^{j2\pi\frac{d_n}{\lambda}(\sin\theta_k+\sin\psi)} s_k +\boldsymbol{w}
\notag                                                                                                                                                        \\
& = \boldsymbol{B}^{\text{T}} \boldsymbol{A}(\boldsymbol{\theta},\boldsymbol{d})\boldsymbol{s} +\boldsymbol{w},
\label{model}
\end{align}
where $\boldsymbol{w} \triangleq \big[w(0), w(T),\dots,w[(M-1)T]\big]^{\text{T}}\in\mathbb{C}^{M\times 1}$, and we have $s_k(mT)\approx s_k((m+1)T) \approx s_k$ ($m=0,\dots,M-2$) with the narrowband assumption. The signal vector is defined as $\boldsymbol{s}\triangleq \begin{bmatrix}
s_0,s_1,\dots ,s_{K-1}
\end{bmatrix}^{\text{T}}\in\mathbb{C}^{K\times 1}$. We define a steering matrix with the direction  $\boldsymbol{\theta}\triangleq [\theta_0, \theta_1,\dots,\theta_{K-1}]^{\text{T}}\in\mathbb{R}^{K\times 1} $ and the position $\boldsymbol{d}$  as
\begin{align}
\boldsymbol{A}(\boldsymbol{\theta},\boldsymbol{d})\triangleq \begin{bmatrix}
\boldsymbol{a}(\theta_0,\boldsymbol{d}), \boldsymbol{a}(\theta_1, \boldsymbol{d}), \dots, \boldsymbol{a}(\theta_{K-1}, \boldsymbol{d})
\end{bmatrix}\in\mathbb{C}^{N\times K},
\end{align}
where the steering vector is defined as
\begin{align}
\boldsymbol{a}(\theta_k,\boldsymbol{d}) & \triangleq e^{j2\pi\frac{\boldsymbol{d}}{\lambda}\sin\theta_k}\in\mathbb{C}^{N\times 1} \\
& =\begin{bmatrix}
e^{j2\pi\frac{d_0}{\lambda}\sin\theta_k}, e^{j2\pi\frac{d_1}{\lambda}\sin\theta_k},\dots, e^{j2\pi\frac{d_{N-1}}{\lambda}\sin\theta_k}
\end{bmatrix}^{\text{T}}.\notag
\end{align}
In (\ref{model}), we also define a measurement matrix as
\begin{align}
\boldsymbol{B}\triangleq \begin{bmatrix}
\boldsymbol{b}(0), \boldsymbol{b}(1),\dots,\boldsymbol{b}(M-1)
\end{bmatrix}\in\mathbb{C}^{N\times M},
\end{align}
where we have
\begin{align}
& \boldsymbol{b}(m)\triangleq \boldsymbol{a}(\psi, \boldsymbol{d})\odot \boldsymbol{c}(m),
\\
& \boldsymbol{c}(m)\triangleq \begin{bmatrix}
A_{0,m}e^{j\phi_{0,m}},
A_{1,m}e^{j\phi_{1,m}},
\dots,
A_{N-1,m}e^{j\phi_{N-1,m}}
\end{bmatrix}^{\text{T}}\label{eq7}.
\end{align}

Therefore, according to the system model in (\ref{model}), the DOA estimation problem using UAV swarm and RIS can be described as the estimation of both the DOA $\boldsymbol{\theta}$ and the signal $\boldsymbol{s}$  from the received signal $\boldsymbol{r}$, where both the measurement matrix $\boldsymbol{B}$ and the position vector $\boldsymbol{d}$ are known.

However, from the practical perspective, these UAV positions cannot be known exactly, and position perturbations must also be considered. We can formulate the system model by introducing the position perturbation. For the $n$-th UAV, the actual position can be expressed as $d_n=\bar{d}_n+\tilde{d}_{n}$, where $\tilde{d}_{n}$ is the position perturbation, and $\bar{d}_n$ is an expected position.

Then, the system model in (\ref{model}) can be rewritten as
\begin{align}
\boldsymbol{r} & = \boldsymbol{B}^{\text{T}}
\text{diag}\{\boldsymbol{a}(\psi, \tilde{\boldsymbol{d}})
\}
\left(\boldsymbol{A}(\boldsymbol{\theta},\bar{\boldsymbol{d}}) \odot \boldsymbol{A}(\boldsymbol{\theta},\tilde{\boldsymbol{d}})\right)\boldsymbol{s} +\boldsymbol{w},\label{eq8}
\end{align}
where the position perturbation vector is defined as $\tilde{\boldsymbol{d}}\triangleq [\tilde{d}_0,\tilde{d}_1,\dots, \tilde{d}_{N-1}]^{\text{T}}$, and the expected position vector is defined as $\bar{\boldsymbol{d}}\triangleq [\bar{d}_0,\bar{d}_1,\dots, \bar{d}_{N-1}]^{\text{T}}$. Additionally, we consider a ULA system, and thus the expected position of the $n$-th UAV is $\bar{d}_n=n\Delta d$, where $\Delta d$ is the distance between the adjacent UAVs, and can be chosen as half of the wavelength $\Delta d=\lambda/2$.

Finally, the DOA estimation problem using the UAV swarm and RIS is expressed as (\ref{eq8}), where the received signal $\boldsymbol{r}$ and the measurement matrix $\boldsymbol{B}$ are known. We try to estimate the DOA $\boldsymbol{\theta}$ with the unknown parameters including the position perturbation $\tilde{\boldsymbol{d}}$, the number of target signals $K$, and the signal $\boldsymbol{s}$.

\section{Super-Resolution DOA Estimation With Position Perturbation}\label{sec3}

To estimate the DOA from the received signal with position perturbation, we propose a super-resolution DOA estimation method by exploiting the signal's sparsity in the spatial domain. The proposed method includes two steps, i.e., the perturbation estimation step and the DOA estimation step.

In the perturbation estimation step, a gradient descent method is formulated. In the DOA estimation step, an atomic norm-based method is proposed, so the proposed method is named as the \emph{A}tomic norm-based \emph{D}OA estimation with \emph{P}osition \emph{P}erturbation (ADPP) method. More details about the proposed ADPP method are given in Algorithm~\ref{adpp}, and will be elaborated in the following subsections.

\begin{algorithm}[t]
	\caption{ADPP Method} \label{adpp}
	\begin{algorithmic}[1]
		\STATE  \emph{Input:} The received signal $\boldsymbol{r}$, the measurement matrix $\boldsymbol{B}$, the expected UAV position $\bar{\boldsymbol{d}}$, the direction of RIS $\psi$,  the maximum number of iterations $Q_1$, and the stop parameters $\epsilon_1$ and $\epsilon_2$.
		\STATE \emph{Initialization:} The estimated position perturbation $\hat{\boldsymbol{d}}=\boldsymbol{0}\in\mathbb{R}^{N\times 1}$, and $q_1=0$.
		\WHILE{$q_1<Q_1$}
		\STATE  The atomic norm-based method is used to estimate the DOA $\hat{\boldsymbol{\theta}}_{q_1}$ during the $q_1$-th iteration, where the estimated position perturbation $\hat{\boldsymbol{d}}_{q_1-1}$ during the $(q_1-1)$-th iteration is used.
		\STATE The  gradient descent method is used to estimate the position perturbation $\hat{\boldsymbol{d}}_{q_1}$, and refine the estimated direction $\hat{\boldsymbol{\theta}}_{q_1}$ during the $q_1$-th iteration, where the estimated DOA $\hat{\boldsymbol{\theta}}_{q_1}$ during the $q_1$-th iteration is used.
		\STATE $q_1\leftarrow q_1+1$.
		\ENDWHILE
		\STATE \emph{Output:} The estimated position perturbation $\hat{\boldsymbol{d}}$, and the estimated DOA $\hat{\boldsymbol{\theta}}$.
	\end{algorithmic}
\end{algorithm}

\subsection{The Atomic Norm-Based Super-Resolution DOA Estimation Method}
\subsubsection{The Realization of DOA Estimation Method}
To estimation the DOA of the received signal, we can exploit the signal sparsity in the spatial domain. The traditional CS-based methods must discretize the spatial domain into grids, and a dictionary matrix is formulated from these discretized grids. Then, a sparse reconstruction method is used in the scenario using the dictionary matrix. Usually, the CS-based techniques can be used for sparse reconstruction. However, these CS-based methods will introduce the \emph{off-grid} problem in the sparse reconstruction processes, where these signals cannot be precisely at these grids. Therefore, atomic norm-based methods have been proposed for the sparse reconstruction without grids and can avoid the \emph{off-grid} problem.

In the DOA estimation problem using the UAV swarm and RIS, the existing atomic norm-based methods cannot be used directly, so we propose a new atomic norm method. Additionally we give a solution of the position perturbation in the sparse reconstruction process.

First, we define a new type of atomic norm by introducing the position perturbation vector $\tilde{\boldsymbol{d}}$, and the atomic norm of $\boldsymbol{x}$ is defined as
\begin{align}
\|\boldsymbol{x}\|_{\mathcal{A}}\triangleq & \inf \bigg\{
\sum_i c_i: \boldsymbol{x}=\sum_i c_i e^{j\phi_i}\boldsymbol{a}(\theta_i,\tilde{\boldsymbol{d}})\odot \boldsymbol{a}(\theta_i,\bar{\boldsymbol{d}}), \notag \\
& \qquad c_i>0, \phi_i\in[0,2\pi),\theta_i\in\left(-\frac{\pi}{2},\frac{\pi}{2}\right]
\bigg\},\label{eq10}
\end{align}
where the non-negative coefficient $c_i\in\mathbb{R}$ describes the atomic decomposition and $\phi_i$ is the corresponding phase. The atomic element is $\boldsymbol{a}(\theta_i,\tilde{\boldsymbol{d}})\odot \boldsymbol{a}(\theta_i,\bar{\boldsymbol{d}})$, where the position perturbation $\tilde{\boldsymbol{d}}$ is considered, so it is different from the existing ANM-based estimation methods, where only an atomic element $\boldsymbol{a}(\theta_i,\bar{\boldsymbol{d}})$ is used. In (\ref{eq10}), we try to find a decomposition of $\boldsymbol{x}$, which is sparse in the domain constructed by atoms $\boldsymbol{a}(\theta_i,\tilde{\boldsymbol{d}})\odot \boldsymbol{a}(\theta_i,\bar{\boldsymbol{d}})$.

Second, based on the definition of atomic norm,  we try to reconstruct a sparse signal from the received signal $\boldsymbol{r}$, and can be formulated as the following reconstruction problem
\begin{align}
\min_{\boldsymbol{x}} \frac{1}{2}\|\boldsymbol{r}-\boldsymbol{B}^{\text{T}}\text{diag}\{\boldsymbol{a}(\psi, \tilde{\boldsymbol{d}})\}\boldsymbol{x}\|^2_2+\beta \|\boldsymbol{x}\|_{\mathcal{A}},\label{eq23}
\end{align}
where $\beta$ is used to control the balance between the sparsity and the reconstruction performance. The second term $\|\boldsymbol{x}\|_{\mathcal{A}}$ describes the sparsity of the signal $\boldsymbol{x}$. To solve the optimization problem (\ref{eq23}), we have the following proposition:
\begin{theorem}\label{pr1}
	The optimization problem (\ref{eq23}) can be expressed as a SDP problem
	\begin{align}
	\min_{\boldsymbol{W},\boldsymbol{h}} & \quad \left[\text{diag}\left\{\boldsymbol{a}^{*}(\psi, \tilde{\boldsymbol{d}})\right\}\boldsymbol{B}^{\text{*}}\boldsymbol{r}-\boldsymbol{h}\right]^{\text{H}}\bigg[\text{diag}\left\{\boldsymbol{a}^{*}(\psi, \tilde{\boldsymbol{d}})\right\}\boldsymbol{B}^{\text{*}}\boldsymbol{B}^{\text{T}}\notag \\
	& \qquad \text{diag}\left\{\boldsymbol{a}(\psi, \tilde{\boldsymbol{d}})\right\}\bigg]^{-1} \left[\text{diag}\left\{\boldsymbol{a}^{*}(\psi, \tilde{\boldsymbol{d}})\right\}\boldsymbol{B}^{\text{*}}\boldsymbol{r}-\boldsymbol{h}\right]\notag                                                           \\
	\text{s.t.}                          & \quad \begin{bmatrix}
	\boldsymbol{W}               & \boldsymbol{Th} \\
	(\boldsymbol{Th})^{\text{H}} & t
	\end{bmatrix}\succeq 0,\label{eq12}                                                                                                                                                                                                                                                 \\
	& \quad \boldsymbol{W} \in\mathbb{C}^{N\times N} \text{ is a Hermitian matrix},\notag                                                                                                                                                                                                                    \\
	& \quad  \text{Tr}\{\boldsymbol{W}\} = \beta^2/t,\notag                                                                                                                                                                                                                                                  \\
	& \quad \sum_{n} W_{n,n+\nu } = 0, \nu \neq 0,\notag
	\end{align}
	where $\boldsymbol{T}\in\mathbb{C}^{N\times N}$ is a transformation matrix, and $t$ is a hyperparameter.
\end{theorem}
The proof of Proposition~\ref{pr1} is given in Appendix~\ref{apd1}.

The next problem is to determine the transformation matrix $\boldsymbol{T}$. Discretize the spatial angle into $\Gamma$ grids with the separation spacing being $\Delta \alpha$, and the $\gamma$-th grid is given by $\gamma\Delta\alpha$. We can formulate following matrices
\begin{align}
\boldsymbol{\Xi}         & \triangleq
\begin{bmatrix}
\boldsymbol{a}(0,\bar{\boldsymbol{d}}), \boldsymbol{a}(\Delta\alpha,\bar{\boldsymbol{d}}),\dots, \boldsymbol{a}((\Gamma-1)\Delta\alpha,\bar{\boldsymbol{d}})
\end{bmatrix},\label{eq33} \\
\tilde{\boldsymbol{\Xi}} & \triangleq
\begin{bmatrix}
\boldsymbol{a}(0,\bar{\boldsymbol{d}}+\tilde{\boldsymbol{d}}),\dots, \boldsymbol{a}((\Gamma-1)\Delta\alpha,\bar{\boldsymbol{d}}+\tilde{\boldsymbol{d}})
\end{bmatrix}.\label{eq34}
\end{align}
With the transformation matrix, we have the following equation
\begin{align}
\boldsymbol{T}^{\text{H}}\boldsymbol{\Xi}=\tilde{\boldsymbol{\Xi}}.
\end{align}
With the vectorization operation, we have
\begin{align}
\text{vec}\left\{\tilde{\boldsymbol{\Xi}}\right\} & = \text{vec}\left\{\boldsymbol{T}^{\text{H}}\boldsymbol{\Xi}\right\} = \left(\boldsymbol{\Xi}^{\text{T}}\otimes \boldsymbol{I}_{N}\right)\text{vec}\left\{\boldsymbol{T}^{\text{H}}\right\},\label{eq36}
\end{align}
where $\boldsymbol{I}_N\in\mathbb{R}^{N\times N}$ is an  identity matrix. For a matrix $\boldsymbol{A}$ with the entry at the $m$-th row and $n$-th column being $a_{m,n}$ ($m=0,1,\dots, M-1$ and $n=0,1,\dots,N-1$), the vectorization operation is defined as
\begin{align}
\text{vec}\{\boldsymbol{A}\}\triangleq & \big[
a_{0,0}, a_{1, 0},\dots, a_{M-1,0}, a_{0,1},\dots, a_{M-1,1},\notag \\
& \qquad \dots, a_{M-1,N-1}
\big]^{\text{T}}.\label{vec}
\end{align}

Then, the vectorized transformation matrix can be estimated as
\begin{align}
\text{vec}\left\{\hat{\boldsymbol{T}}^{\text{H}}\right\} = \left(\boldsymbol{\Xi}^{\text{T}}\otimes \boldsymbol{I}_{N}\right)^{\dagger} \text{vec}\left\{\tilde{\boldsymbol{\Xi}}\right\}.\label{eq38}
\end{align}

Using the inverse transformation
of the vectorization operation in (\ref{vec}), the transformation matrix can be estimated as $\hat{\boldsymbol{T}}$ from $\text{vec}\left\{\hat{\boldsymbol{T}}^{\text{H}}\right\}$.

\begin{algorithm}[t]
	\caption{Atomic Norm-Based DOA Estimation Method} \label{anm}
	\begin{algorithmic}[1]
		\STATE  \emph{Input:} The received signal $\boldsymbol{r}$, the measurement matrix $\boldsymbol{B}$, the estimated position perturbation $\tilde{\boldsymbol{d}}$, the expected UAV's position $\bar{\boldsymbol{d}}$, the direction of RIS $\psi$, and the hyperparameter $t$.
		\STATE Discretize the spatial domain into $\Gamma$ grids, and formulate matrices as  (\ref{eq33}) and (\ref{eq34}) using the estimated perturbation $\tilde{\boldsymbol{d}}$.
		\STATE Obtain the transformation matrix $\boldsymbol{T}$ from (\ref{eq36}).
		\STATE Formulate the SDP optimization problem as (\ref{eq12}).
		\STATE Use the CVX toolbox to obtain $\boldsymbol{h}$ from (\ref{eq12}).\label{step5}
		\STATE Find the peak values from (\ref{eq32}), and the positions of the peak values are corresponding to the estimated DOA.
		\STATE \emph{Output:} The estimated DOA $\hat{\boldsymbol{\theta}}$.
	\end{algorithmic}
\end{algorithm}

Finally, the DOA estimation method are given in Algorithm~\ref{anm}. The atomic norm-based super-resolution method for the DOA estimation using the UAV swarm and RIS can be obtained as the optimization problem in (\ref{eq12}), where the transformation matrix is given in (\ref{eq38}). The vector $\boldsymbol{h}$ is obtained from (\ref{eq12}) using the convex optimization toolbox, such as CVX toolbox in MATLAB~\cite{cvx,gb08}.  According to the dual constraint in (\ref{eq32}), the vector $\boldsymbol{h}$ must satisfy the following condition
\begin{align}\label{eq39}
f(\theta)\triangleq \left|
\boldsymbol{h}^{\text{H}} \boldsymbol{T}^{\text{H}}\boldsymbol{a}(\theta, \bar{\boldsymbol{d}})\right|\leq \beta, \forall \theta\in\left[-\frac{\pi}{2},\frac{\pi}{2}\right),
\end{align}
and the  peak value of the expression can be obtained and the position  is corresponding to the estimated DOA $\hat{\boldsymbol{\theta}}$.

\subsubsection{The Simulation Validation}
In the practical RIS system, only a finite number of phase shifts can be chosen to control the reflection angle using RIS. Hence, in the measurement $\boldsymbol{B}$, the amplitude is the same, i.e., $A_{0,m}=A_{1,m}=\dots=A_{N-1,m}$, and the phase is in the discretized set, i.e, $\phi_{n,m}\in \aleph=\left\{\phi_1,\phi_2,\dots, \phi_{|\aleph|}\right\}$, where $|\aleph|$ denotes the cardinality of the set $\aleph$. In this paper, only 2 phases are controlled by RIS, i.e., $|\aleph| = 2$, and $\aleph=\{0, \pi\}$, so the entry of $\boldsymbol{c}(m)$ in (\ref{eq8}) is chosen as $1$ or $-1$ randomly.

Here, we give some simulation results using the proposed DOA estimation method. The number of UAVs is $N=32$, the number of measurements is $M=32$, and the distance between adjacent UAVs is $\lambda/2$.
With the SNR being $30$ dB, the direction between the UAV swarm and the central UAV is $\psi=0$. The measurement matrix is a random matrix.

\begin{figure}
	\centering
	\includegraphics[width=3.8in]{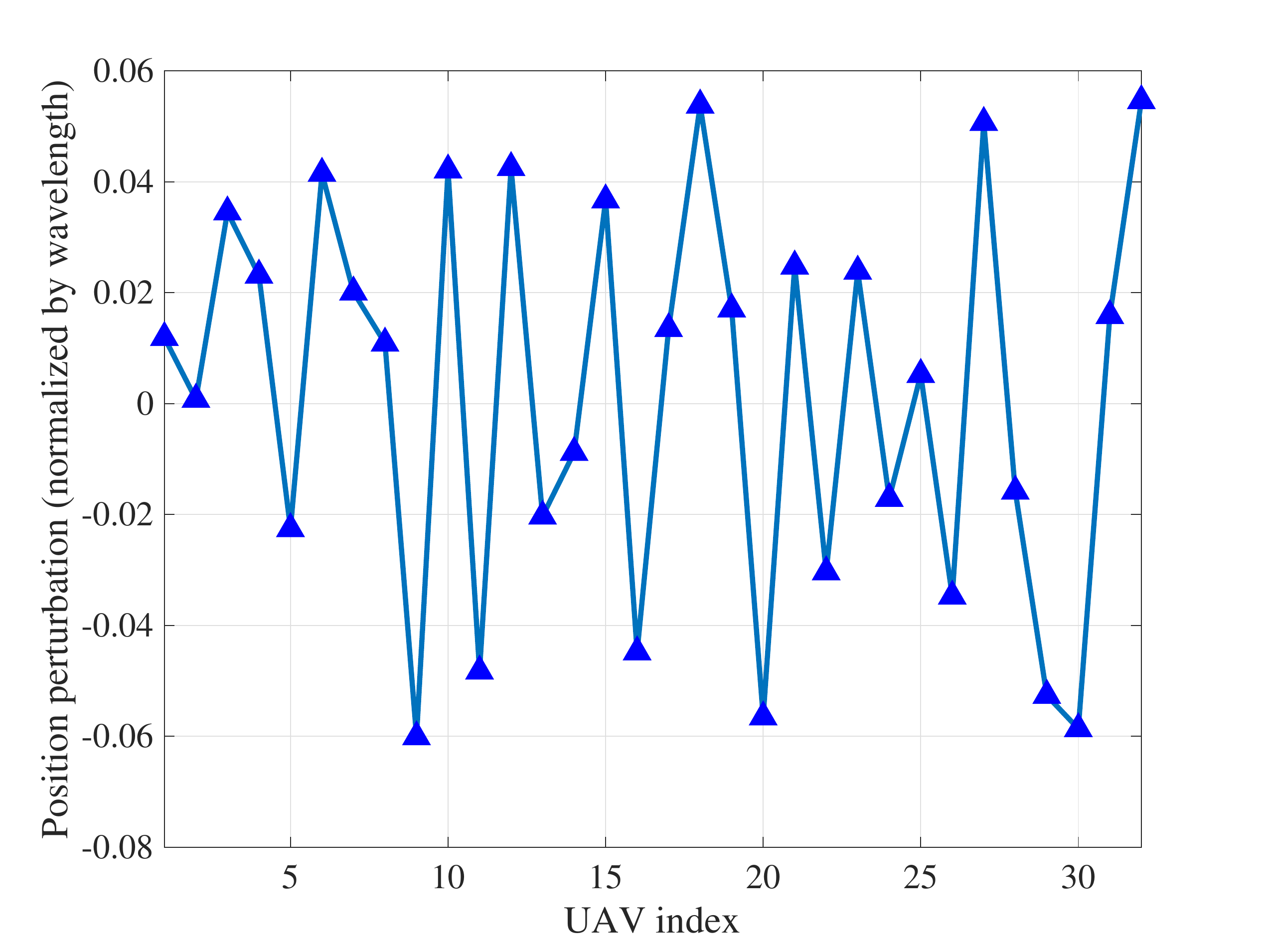}
	\caption{The position perturbations (normalized by wavelength).}
	\label{perturbation}
\end{figure}

The position perturbation of the UAV follows a uniform distribution, i.e., $\tilde{d}_n\in\left(-\frac{\lambda}{16}, \frac{\lambda}{16}\right]$, and Fig.~\ref{perturbation} gives an example normalized by wavelength. After obtaining the transformation matrix, the vector $\boldsymbol{h}$ can be solved from (\ref{eq12}), where we can choose $t=\beta^2$, so $\text{Tr}\{\boldsymbol{W}\} = 1$ and the Hermitian matrix constraint will be $\begin{bmatrix}
\boldsymbol{W},               & \boldsymbol{Th} \\
(\boldsymbol{Th})^{\text{H}}, & \beta^2
\end{bmatrix}\succeq 0.$

\begin{figure}
	\centering
	\includegraphics[width=3.8in]{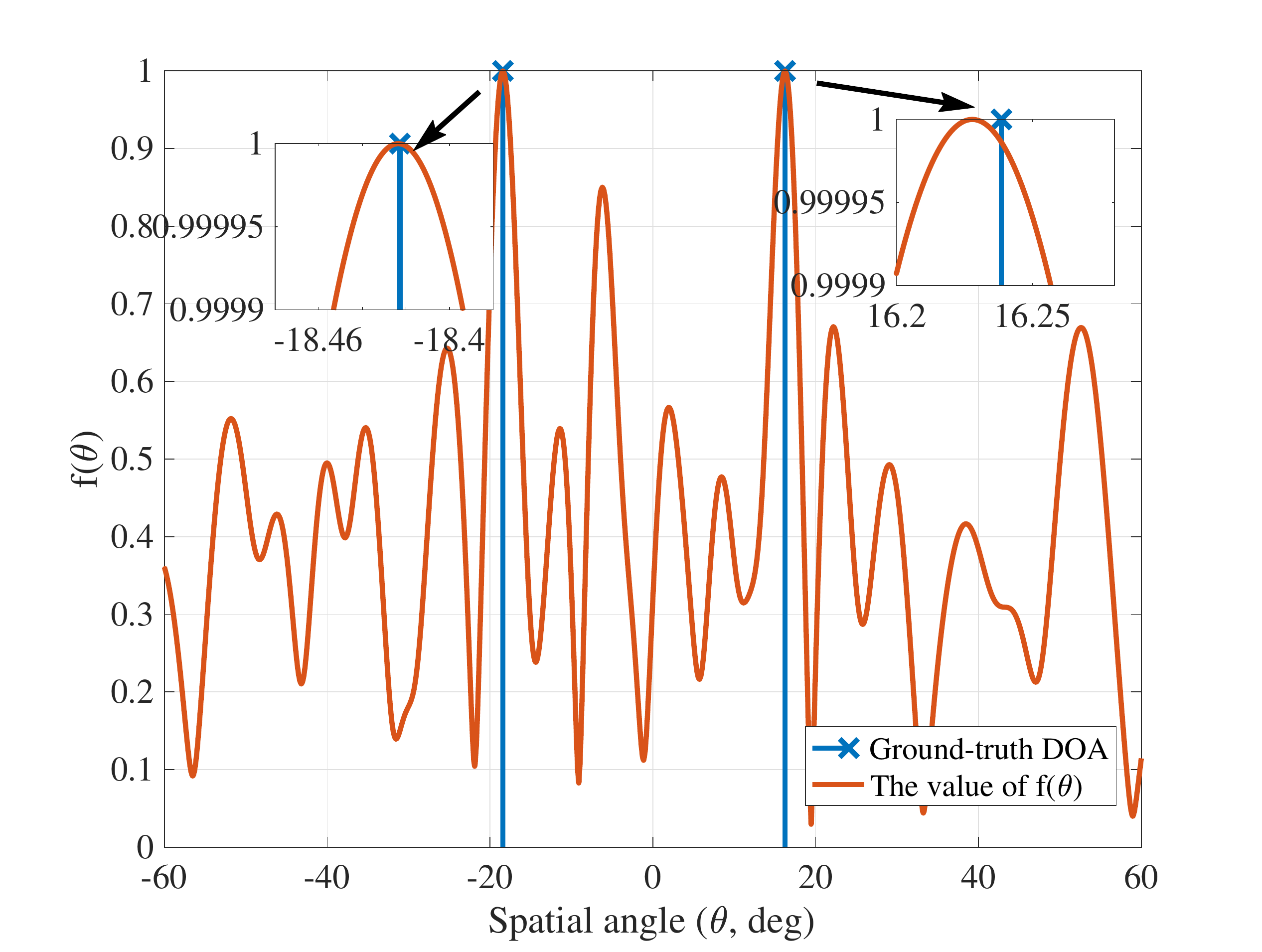}
	\caption{The values of $f(\theta)$.}
	\label{f30dB}
\end{figure}

Then, the values of the function $f(\theta)$ can be obtained and shown in Fig.~\ref{f30dB}. We consider $2$ targets with the DOA being $\ang{-18.4228}$ and $\ang{16.2385}$. The hyperparameter $t=\beta^2=500$. As shown in Fig.~\ref{f30dB}, the peak values of $f(\theta)$ ared corresponding to the ground-truth DOAs, so the estimated DOAs are $\ang{-18.424}$ and $\ang{16.228}$, respectively. The root mean square error (RMSE) for the DOA estimation is $\ang{0.0075}$. Therefore, the DOA can be estimated with high precision.

\begin{figure}
	\centering
	\includegraphics[width=3.8in]{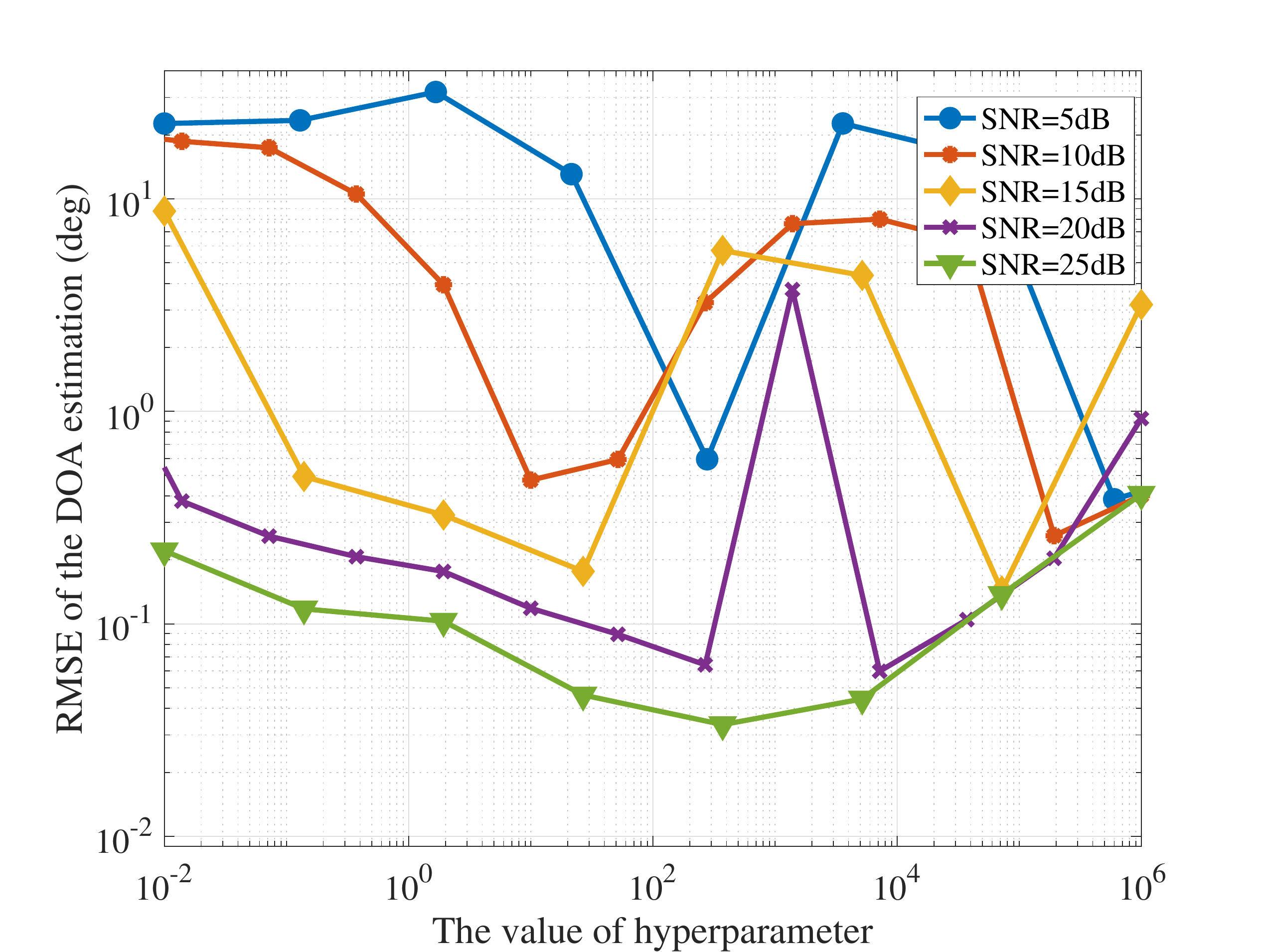}
	\caption{The RMSE of DOA estimation with different hyperparameter values.}
	\label{hyper}
\end{figure}

In the atomic norm-based DOA estimation method, the hyperparameter $t$ is essential and determines the DOA estimation performance. With $t=\beta^2$, we show the RMSE of the DOA estimation with different hyperparameter values in Fig.~\ref{hyper}. As shown in this figure, when the SNR is lesser than $10$ dB, we can choose the hyperparameter as $t=200$. When the SNR is great than $10$ dB, we can choose the hyperparameter as $t=10$. Then, the better DOA estimation performance can be achieved by the atomic norm-based method. For the adaptive application, the hyperparameter can be obtained from the following expression
\begin{align}
t=\begin{cases}
e^{-0.5991\text{SNR}+8.294}, & \text{SNR}\leq 10 \text{ dB}\\
e^{0.2593\text{SNR}-0.2889}, & \text{SNR}> 10 \text{ dB}
\end{cases},
\end{align}
and it is an empirical expression based on simulation results.

\subsection{The Gradient Descent Method for Perturbation Estimation}
\subsubsection{The Perturbation Estimation Method}
Since the targets are sparse in the spatial domain, the atomic norm-based method is proposed to reconstruct the sparse signal with high resolution. However, in the DOA estimation problem using UAV swarm and RIS, the unknown position perturbation, as shown in (\ref{eq8}), has a significant effect on the DOA estimation performance. Additionally, the position perturbation causes phase error in the system model, making the estimation problem even more challenging.

Therefore, inspired by the nonconvex optimization and gradient descent method in~\cite{9410615,9398573}, we propose a gradient descent method for the perturbation estimation, i.e., to estimate $\tilde{\boldsymbol{d}}$ from the received signal $\boldsymbol{r}$ in (\ref{eq8}).

\begin{algorithm}%[t]
	\caption{Position Perturbation Estimation Method} \label{alg3}
	\begin{algorithmic}[1]
		\STATE  \emph{Input:} The received signal $\boldsymbol{r}$, the measurement matrix $\boldsymbol{B}$, the estimated signal direction $\hat{\boldsymbol{\theta}}$, the expected UAV's position $\bar{\boldsymbol{d}}$, the direction of RIS $\psi$, the number of signals $K$, the step size $\varrho_1$, the step size $\varrho_2$, and the maximum number of iterations $Q_2$.
		\STATE \emph{Initialization:} The initial position perturbation $\hat{\boldsymbol{d}}=\boldsymbol{0}\in\mathbb{R}^{N\times 1}$, $\epsilon=10^{-3}$, and $q_2=0$.
		\WHILE{$q_2<Q_2$}
		\STATE \label{alg3step4} Obtain the following matrix $\boldsymbol{V}$
		\begin{align}
		\boldsymbol{V} & =\boldsymbol{B}^*\boldsymbol{B}^{\text{T}}\text{diag}\bigg\{[\boldsymbol{A}(\hat{\boldsymbol{\theta}}, \bar{\boldsymbol{d}})\odot \boldsymbol{A}(\hat{\boldsymbol{\theta}},\hat{\boldsymbol{d}}) \odot \boldsymbol{A}(\psi\boldsymbol{1}_K, \hat{\boldsymbol{d}})]\notag \\
		& \qquad \text{diag}\big\{\frac{j2\pi}{\lambda} (\sin\psi+\sin\hat{\boldsymbol{\theta}})\big\}\boldsymbol{s} \bigg \}.\end{align}
		\STATE \label{alg3step5} Obtain the following matrix $\boldsymbol{G}$
		\begin{align}
		\boldsymbol{G}=\text{diag}\{\boldsymbol{s}\}^{\text{H}}\left[\boldsymbol{A}(\hat{\boldsymbol{\theta}},\hat{\boldsymbol{d}})\odot \boldsymbol{A}(\psi \boldsymbol{1}_K,\hat{\boldsymbol{d}})\odot \boldsymbol{A}(\hat{\boldsymbol{\theta}}, \bar{\boldsymbol{d}})
		\right]^{\text{H}}\boldsymbol{V}.
		\end{align}
		\STATE \label{alg3step6} Obtain the following matrix $\boldsymbol{H}$
		\begin{align}
		\boldsymbol{H} & =\text{diag}\big\{\frac{j2\pi}{\lambda}(\sin\psi+\sin\hat{\boldsymbol{\theta}})\odot \boldsymbol{s}\big\}\bigg[
		\boldsymbol{A}(\hat{\boldsymbol{\theta}},\bar{\boldsymbol{d}})\notag                                                                                           \\
		& \qquad \odot \boldsymbol{A}(\psi \boldsymbol{1}_K, \hat{\boldsymbol{d}})\odot \boldsymbol{A}(\hat{\boldsymbol{\theta}}, \hat{\boldsymbol{d}})
		\bigg]^{\text{T}}\text{diag}\{\boldsymbol{r}^{\text{H}}\boldsymbol{B}^{\text{T}}\}.
		\end{align}
		\STATE \label{alg3step7} The gradient vector is defined as $\nabla \hat{\boldsymbol{d}}\triangleq \left[\nabla \hat{d}_0, \nabla \hat{d}_1,\dots, \nabla \hat{d}_{N-1}\right]^{\text{T}}$, where the $n$-th entry is obtained as
		\begin{align}
		\nabla \hat{d}_n=2\mathcal{R}\left\{\sum_{k=0}^{K-1}(G_{k,n}-H_{k,n})\right\},\label{eq52}
		\end{align}
		where $G_{k,n}$ and $H_{k,n}$ are the entries of $\boldsymbol{G}$ and $\boldsymbol{H}$ at the $k$-th row and $n$-th column, respectively.
		\STATE \label{alg3step8} Update the estimated position perturbation as
		\begin{align}
		\hat{\boldsymbol{d}}\leftarrow \hat{\boldsymbol{d}}-\varrho_1 \nabla \hat{\boldsymbol{d}}.\label{eq55}
		\end{align}
		\STATE \label{alg3step9} Obtain the gradient vector $	\frac{\partial \eta (\tilde{\boldsymbol{d}},\boldsymbol{\theta})}{\partial \boldsymbol{\theta}}$, and tune the estimated direction as
		\begin{align}
		\hat{\boldsymbol{\theta}}\leftarrow \hat{\boldsymbol{\theta}}-\varrho_2 \nabla \hat{\boldsymbol{\theta}}.
		\end{align}
		\STATE \label{alg3step10} Obtain the estimated signal during the $q_2$-th iteration as
		\begin{align}
		\hat{\boldsymbol{r}}_{q_2}=\boldsymbol{B}^{\text{T}}\text{diag}\{\boldsymbol{a}(\psi, \hat{\boldsymbol{d}})\}[\boldsymbol{a}(\hat{\boldsymbol{\theta}},\bar{\boldsymbol{d}})\odot\boldsymbol{a}(\hat{\boldsymbol{\theta}},\hat{\boldsymbol{d}})]\boldsymbol{s}.
		\end{align}
		\IF{$\|\boldsymbol{r}-\hat{\boldsymbol{r}}_{q_2}\|^2_2/\|\boldsymbol{r}_{q_2-1}-\hat{\boldsymbol{r}}\|^2_2\leq \epsilon$}
		\STATE Break.
		\ENDIF
		\STATE $q_2\leftarrow q_2+1$.
		\ENDWHILE
		\STATE \emph{Output:} The estimated position perturbation $\hat{\boldsymbol{d}}$.
	\end{algorithmic}
\end{algorithm}

The algorithm of the position perturbation estimation is given in Algorithm~\ref{alg3}, where $\boldsymbol{1}_K\in\mathbb{R}^{K\times 1}$ denotes a vector with all entries being $1$. In the gradient descent method, the estimated position perturbation is updated by the gradient vector, as shown in (\ref{eq55}), where the most critical part is to obtain the gradient vector. The following contents are about how to get the gradient vector.

First, the system model in (\ref{eq8}) can be rewritten as
\begin{align}
\boldsymbol{r} = \boldsymbol{B}^{\text{T}} \sum^{K-1}_{k=0} \text{diag}\{\boldsymbol{e}(\theta_k)\}\boldsymbol{a}(\theta_k) s_k+\boldsymbol{w},
\end{align}
where we define the $n$-th entry of $\boldsymbol{e}(\theta_k)$ as
\begin{align}
e_n(\theta_k) = e^{j2\pi\frac{\tilde{d}_n}{\lambda}(\sin\theta_k+\sin\psi)}.
\end{align}

To estimate the position perturbation, we can formulate the following optimization problem
\begin{align}
\{\hat{\boldsymbol{d}},\hat{\boldsymbol{\theta}}\}=\arg\min_{\tilde{\boldsymbol{d}},\boldsymbol{\theta}}\, \eta(\tilde{\boldsymbol{d}}, \boldsymbol{\theta}),\label{eq44}
\end{align}
where $\hat{\boldsymbol{d}}$ denotes the estimated position perturbation, and $\hat{\boldsymbol{\theta}}$ is the tuned DOA, and
\begin{align}
\eta(\tilde{\boldsymbol{d}}, \boldsymbol{\theta})\triangleq \bigg\|\boldsymbol{r}-\boldsymbol{B}^{\text{T}} \sum^{K-1}_{k=0} \text{diag}\{\boldsymbol{e}(\theta_k)\}\boldsymbol{a}(\theta_k) s_k\bigg\|^2_2.\label{eq50}
\end{align}

Then, the optimization problem (\ref{eq44}) can be simplified as
\begin{align}
\{\hat{\boldsymbol{d}},\boldsymbol{\theta}\} & =\arg\min_{\tilde{\boldsymbol{d}},\boldsymbol{\theta}}\, \eta(\tilde{\boldsymbol{d}},\boldsymbol{\theta})                                            \\
& =  \arg\min_{\tilde{\boldsymbol{d}},\boldsymbol{\theta}} -
\boldsymbol{r}^{\text{H}}\left(\boldsymbol{B}^{\text{T}} \sum^{K-1}_{k=0} \text{diag}\{\boldsymbol{e}(\theta_k)\}\boldsymbol{a}(\theta_k) s_k\right)\notag                                          \\
& \qquad -\left(\boldsymbol{B}^{\text{T}} \sum^{K-1}_{k=0} \text{diag}\{\boldsymbol{e}(\theta_k)\}\boldsymbol{a}(\theta_k) s_k\right)^{\text{H}}\notag \\
& \qquad
\left(\boldsymbol{r}-\boldsymbol{B}^{\text{T}} \sum^{K-1}_{k=0} \text{diag}\{\boldsymbol{e}(\theta_k)\}\boldsymbol{a}(\theta_k) s_k\right)\notag
\end{align}

Then, we have
\begin{align}
\frac{\partial \eta(\tilde{\boldsymbol{d}},\boldsymbol{\theta})}{\partial \tilde{d}_n} & =
\sum^{K-1}_{k=0} 2\mathcal{R}\Bigg\{-s_k \left[\boldsymbol{r}^{\text{H}}\boldsymbol{B}^{\text{T}} \right]_n a_n(\theta_k) \frac{\partial e_n(\theta_k)  }{\partial \tilde{d}_n}\notag \\
& \qquad + \left( s^*_k
\boldsymbol{a}^{\text{H}} (\theta_k)
\text{diag}\{\boldsymbol{e}^{\text{*}}(\theta_k)\}\right) \boldsymbol{B}^{\text{*}} [\boldsymbol{B}^{\text{T}}]_n \notag                                                              \\
& \qquad\bigg(\sum^{K-1}_{k'=0} \frac{\partial
	e_n(\theta_{k'})
}{\partial \tilde{d}_n} a_n(\theta_{k'})  s_{k'} \bigg)\Bigg\},
\end{align}
where we have
\begin{align}
\frac{\partial e_n(\theta_k)}{\partial \tilde{d}_n} = \frac{j2\pi}{\lambda}(\sin\theta_k+\sin\psi) e^{j2\pi\frac{\tilde{d}_n}{\lambda}(\sin\theta_k+\sin\psi)}.
\end{align}
Finally, the gradient vector can be simplified as (\ref{eq52}) in Algorithm~\ref{alg3}.

Moreover, for the DOA estimation, we can use the gradient descent method to refine the estimation result. Hence, the expression of $\frac{\partial \eta (\tilde{\boldsymbol{d}},\boldsymbol{\theta})}{\partial \theta_k}$ is given in (\ref{eq66}). Then, collect these values $\frac{\partial \eta (\tilde{\boldsymbol{d}},\boldsymbol{\theta})}{\partial \theta_k}$ ($k=0,1,\dots,K-1$) into a vector, and we can simplify it as
\begin{align}
\frac{\partial \eta (\tilde{\boldsymbol{d}},\boldsymbol{\theta})}{\partial \boldsymbol{\theta}} & = 2\mathcal{R}\Bigg\{
\frac{j2\pi}{\lambda}\text{diag}\{\cos\boldsymbol{\theta}\odot \boldsymbol{s}\} \boldsymbol{A}^{\text{T}}(\boldsymbol{\theta},\tilde{\boldsymbol{d}}+\bar{\boldsymbol{d}})\notag                                                  \\
& \qquad \qquad \text{diag}\{ (\tilde{\boldsymbol{d}}+\bar{\boldsymbol{d}})\odot \boldsymbol{a}(\psi ,\tilde{\boldsymbol{d}})  \}
\tilde{\boldsymbol{z}}
\Bigg\},
\end{align}
where we have $\tilde{\boldsymbol{z}}\triangleq \boldsymbol{B}\bigg(\boldsymbol{B}^{\text{T}} \sum^{K-1}_{k=0} \text{diag}\{\boldsymbol{e}(\theta_k)\}\boldsymbol{a}(\theta_k) s_k-\boldsymbol{r}\bigg)^*$.

\subsubsection{The Simulation Validation}

\begin{figure}
	\centering
	\includegraphics[width=3.6in]{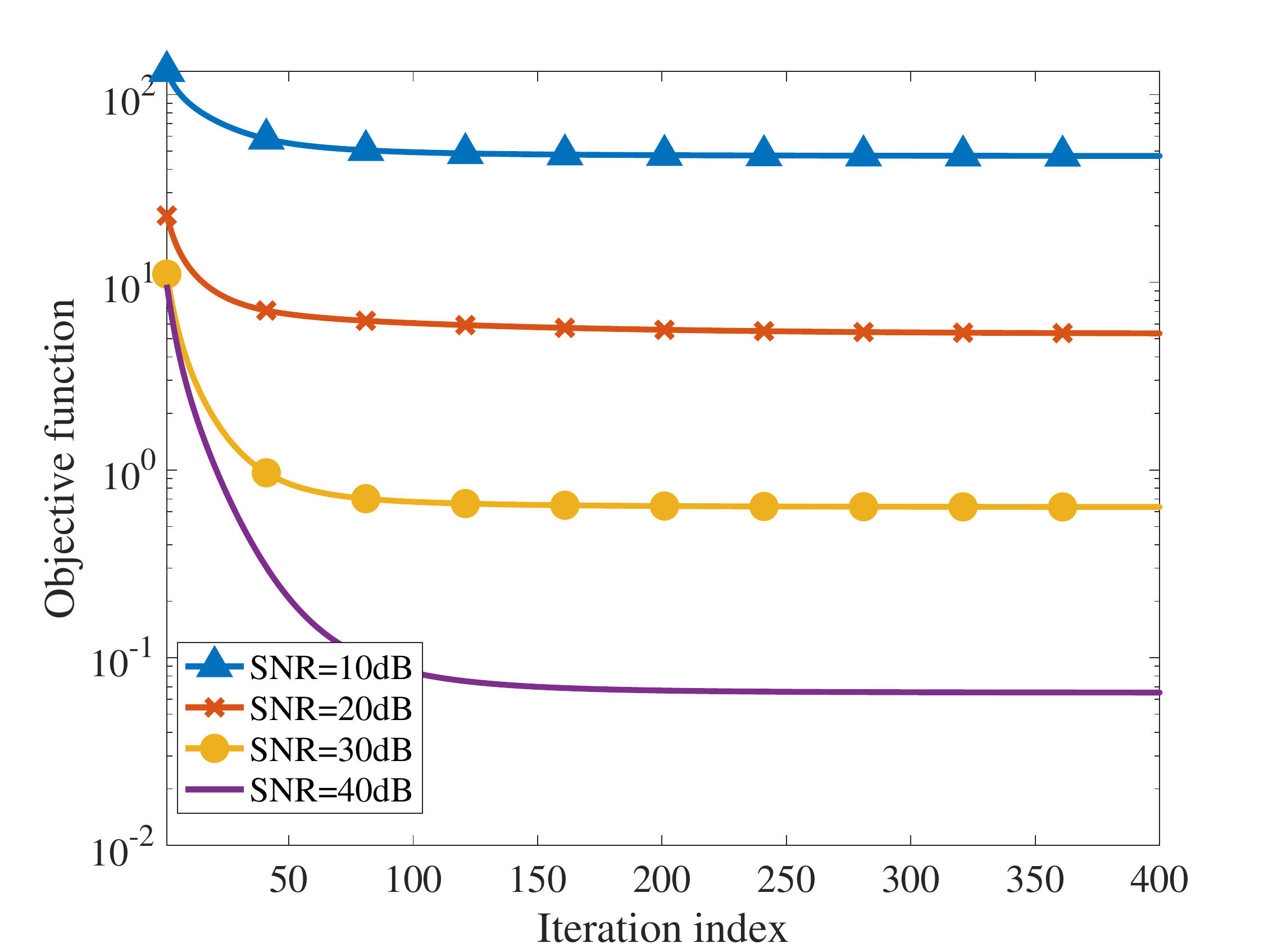}
	\caption{The objective function $\eta(\tilde{\boldsymbol{d}})$.}
	\label{obj}
\end{figure}

\begin{figure}
	\centering
	\includegraphics[width=3.6in]{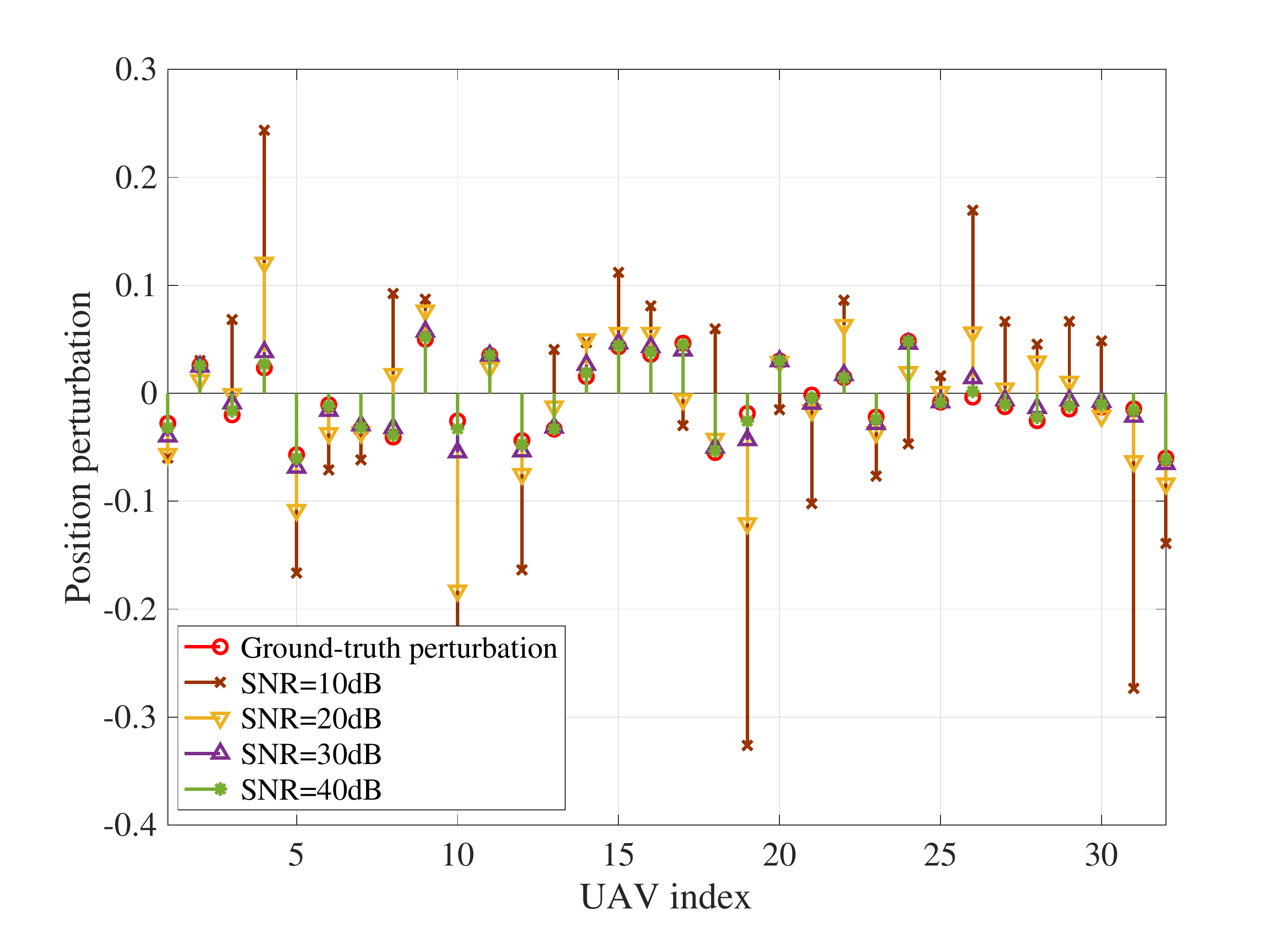}
	\caption{The estimated position perturbation.}
	\label{estPer}
\end{figure}

Using $32$ UAVs, with the knowledge of the signal direction $\boldsymbol{\theta}$, we show the estimated position perturbation using the proposed gradient descent method. The position perturbations follow a uniform distribution in $\left[-\frac{\lambda}{16}, \frac{\lambda}{16}\right)$.

With the SNR being $10$ dB, $20$ dB, $30$ dB and $40$ dB,  the objective function $\eta(\tilde{\boldsymbol{d}})$ in (\ref{eq50}) is shown in Fig.~\ref{obj}. As shown in this figure,  the objective function is stable when the iteration number is greater than $100$. The estimated result is shown in Fig.~\ref{estPer}, and a better estimation performance is achieved with SNR exceeding $30$ dB.

\subsection{Computational Complexity}
The computational complexity of the proposed ADPP method can be estimated from the main steps in Algorithm~\ref{anm} and Algorithm~\ref{alg3}. In Algorithm~\ref{anm}, the SDP problem is solved and  the computational complexity is determined by step~\ref{step5}, which is $\mathcal{O}((N+1)^{6.5})$~\cite{kalantari2020equivalence}. In Algorithm~\ref{alg3}, the computational complexities for step~\ref{alg3step4},  step~\ref{alg3step5}, step~\ref{alg3step6}, step~\ref{alg3step9} and step~\ref{alg3step10} are $\mathcal{O}(N^2(M+K))$, $\mathcal{O}(N^2+NK)$, $\mathcal{O}(N^2(K+1)+K^2N)$, $\mathcal{O}(N^3+KN^2+K^2N)$ and $\mathcal{O}(N^2)$, respectively. Usually, for the DOA estimation problem, we have $N>K$. Therefore, the computational complexity of Algorithm~\ref{alg3} can be simplified as $\mathcal{O}(N^2(N+M))$. Finally, the computational complexity of ADPP method can be obtained as $\mathcal{O}((N+1)^{6.5}+N^2M)$.

	\section{Cram\'{e}r-Rao Bound (CRB)  for the DOA Estimation With Position Perturbation}\label{sec4}
	To show the estimation performance in the DOA estimation problem with position perturbation, these unknown parameters including the DOA and the UAVs position, and we collect all unknown parameters into a vector as
	\begin{align}
		\boldsymbol{\zeta}\triangleq \begin{bmatrix}
			\tilde{d}_0,\tilde{d}_1,\dots, \tilde{d}_{N-1},\theta_0,\dots, \theta_{K-1}
		\end{bmatrix}^{\text{T}}.
	\end{align}
	For the received signal $\boldsymbol{r}$ with AWGN $\boldsymbol{w}$, we have the following probability density function
	\begin{align}
		f(\boldsymbol{r}; \boldsymbol{\zeta}) = \frac{1}{\pi^{M}\det(\boldsymbol{\Sigma})}e^{-(\boldsymbol{r}-\boldsymbol{\mu})^{\text{H}}\boldsymbol{\Sigma}^{-1}(\boldsymbol{r}-\boldsymbol{\mu})},
	\end{align}
	where the mean and the covariance matrix of the received signal are  denoted respectively as
	\begin{align}
		\boldsymbol{\mu}    & =\boldsymbol{B}^{\text{T}} \sum^{K-1}_{k=0} \text{diag}\{\boldsymbol{e}(\theta_k)\}\boldsymbol{a}(\theta_k) s_k, \\
		\boldsymbol{\Sigma} & =\sigma^2_{\text{w}}\boldsymbol{I}.
	\end{align}

	Then, we can obtain the Fisher information matrix $\boldsymbol{F}$~\cite{9449977} as
	\begin{align}
		\boldsymbol{F} = \begin{bmatrix}\boldsymbol{F}_{1,1}, & \boldsymbol{F}_{1,2} \\
			\boldsymbol{F}_{2,1}, & \boldsymbol{F}_{2,2}
		\end{bmatrix},
	\end{align}
	where for $\boldsymbol{F}_{1,1}$, we have
\begin{align}
\boldsymbol{F}_{1,1} & =\mathcal{E}\left\{\frac{\partial \ln f(\boldsymbol{r};\boldsymbol{\zeta})}{\partial \tilde{\boldsymbol{d}}}\frac{\partial \ln f^{\text{T}}(\boldsymbol{r};\boldsymbol{\zeta})}{\partial \tilde{\boldsymbol{d}}}\right\} \\
                     & = \mathcal{E}\bigg\{\left[
(\boldsymbol{r}-\boldsymbol{\mu})^{\text{H}} \boldsymbol{\Sigma}^{-1}\frac{\partial \boldsymbol{\mu}}{\partial \tilde{\boldsymbol{d}}}+[\boldsymbol{\Sigma}^{-1}(\boldsymbol{r}-\boldsymbol{\mu})]^{\text{T}}\frac{\partial \boldsymbol{\mu}^{\text{*}}}{\partial \tilde{\boldsymbol{d}}}
\right]\notag                                                                                                                                                                                                                                   \\
                     & \qquad\qquad\frac{\partial f^{\text{T}}(\boldsymbol{r};\boldsymbol{\zeta})}{\partial \tilde{\boldsymbol{d}}}\bigg\}\notag                                                                                                \\
                     & = \frac{4}{\sigma_{\text{w}}^4}\mathcal{E}\Bigg\{\mathcal{R}\bigg\{
\boldsymbol{\Lambda}^{\text{T}}(\boldsymbol{r}-\boldsymbol{\mu})^{\text{*}} \bigg\}
\mathcal{R}\bigg\{
(\boldsymbol{r}-\boldsymbol{\mu})^{\text{H}} \boldsymbol{\Lambda}\bigg\}\Bigg\}\notag                                                                                                                                                           \\
                     & = \frac{2} {\sigma^2_{\text{w}}} \mathcal{R}\left\{\boldsymbol{\Lambda}^{\text{H}} \boldsymbol{\Lambda}
\right\},\notag
\end{align}
	and we define
	\begin{align}
		\boldsymbol{\Lambda}\triangleq & \boldsymbol{B}^{\text{T}} \sum^{K-1}_{k=0}\bigg[\text{diag}\{ \boldsymbol{a}(\theta_k) \odot e^{j2\pi\frac{\tilde{\boldsymbol{d}}}{\lambda}(\sin\theta_k+\sin\psi)}\}\notag \\
		                               & \quad\quad \frac{s_k j 2\pi(\sin\theta_k+\sin\psi)}{\lambda}\bigg].
	\end{align}
	Similarly, other sub-matrices can be obtained as
	\begin{align}
		\boldsymbol{F}_{1,2} & = \mathcal{E}\left\{\frac{\partial \ln f(\boldsymbol{r};\boldsymbol{\zeta})}{\partial \tilde{\boldsymbol{d}}}\frac{\partial \ln f^{\text{T}}(\boldsymbol{r};\boldsymbol{\zeta})}{\partial \boldsymbol{\theta}}\right\} \\
		                     & = \frac{2} {\sigma^2_w} \mathcal{R}\left\{\boldsymbol{\Lambda}^{\text{H}} \boldsymbol{\Upsilon}
		\right\},\notag                                                                                                                                                                                                                               \\
		\boldsymbol{F}_{2,1} & = \mathcal{E}\left\{\frac{\partial \ln f(\boldsymbol{r};\boldsymbol{\zeta})}{\partial \boldsymbol{\theta}}\frac{\partial \ln f^{\text{T}}(\boldsymbol{r};\boldsymbol{\zeta})}{\partial \tilde{\boldsymbol{d}}}\right\} \\
		                     & = \frac{2} {\sigma^2_w} \mathcal{R}\left\{\boldsymbol{\Upsilon}^{\text{H}} \boldsymbol{\Lambda}
		\right\},\notag                                                                                                                                                                                                                               \\
		\boldsymbol{F}_{2,2} & = \mathcal{E}\left\{\frac{\partial \ln f(\boldsymbol{r};\boldsymbol{\zeta})}{\partial \boldsymbol{\theta}}\frac{\partial \ln f^{\text{T}}(\boldsymbol{r};\boldsymbol{\zeta})}{\partial \boldsymbol{\theta}}\right\}    \\
		                     & = \frac{2} {\sigma^2_w} \mathcal{R}\left\{\boldsymbol{\Upsilon}^{\text{H}} \boldsymbol{\Upsilon}
		\right\},\notag
	\end{align}
	where the definition of $\boldsymbol{\Upsilon}$ is given in (\ref{eq64}).

	\begin{figure*}[!t]
		\normalsize
		\begin{align}
			\boldsymbol{\Upsilon} & \triangleq \frac{\partial \boldsymbol{\mu}}{\partial \boldsymbol{\theta}}  = \boldsymbol{B}^{\text{T}} \sum^{K-1}_{k=0}    s_k\frac{\partial \text{diag}\{\boldsymbol{e}(\theta_k)\}\boldsymbol{a}(\theta_k)}{\partial \boldsymbol{\theta}}\label{eq64} \\
			                      & = \boldsymbol{B}^{\text{T}} \begin{bmatrix}\dots,\underbrace{j\frac{2\pi}{\lambda} s_k
					\cos\theta_k 	e^{j\frac{2\pi}{\lambda} [\boldsymbol{\tilde{d}} (\sin\theta_k+\sin\psi) +\boldsymbol{d} \sin\theta_k]}\odot (\boldsymbol{\tilde{d}}  +\boldsymbol{d} )}_{\text{the $k$-th column}} , \dots\end{bmatrix}.\notag
		\end{align}

		\begin{align}
			\frac{\partial \eta (\boldsymbol{\theta})}{\partial \theta_k}
			 & = 2\mathcal{R}\Bigg\{ \bigg(\boldsymbol{B}^{\text{T}} \sum^{K-1}_{k=0} \text{diag}\{\boldsymbol{e}(\theta_k)\}\boldsymbol{a}(\theta_k) s_k-\boldsymbol{r}\bigg)^{\text{H}}\boldsymbol{B}^{\text{T}}  s_k \frac{\partial   \text{diag}\{\boldsymbol{e}(\theta_k)\}\boldsymbol{a}(\theta_k) }{\partial \theta_k}\Bigg\}\notag \\
			 & = 2\mathcal{R}\Bigg\{ \bigg(\boldsymbol{B}^{\text{T}} \sum^{K-1}_{k=0} \text{diag}\{\boldsymbol{e}(\theta_k)\}\boldsymbol{a}(\theta_k) s_k-\boldsymbol{r}\bigg)^{\text{H}}\boldsymbol{B}^{\text{T}}  s_k \Bigg[\text{diag}\{\boldsymbol{e}(\theta_k)\}\boldsymbol{a}(\theta_k)\odot \begin{pmatrix}
					\vdots                                                               \\
					j\frac{2\pi}{\lambda}\left(\tilde{d}_n+\bar{d}_n \right)\cos\theta_k \\
					\vdots
				\end{pmatrix}\Bigg]
			\Bigg\}.\label{eq66}
		\end{align}

		\hrulefill
		\vspace*{4pt}
	\end{figure*}

	Therefore, the Fisher information matrix $\boldsymbol{F}$ can be simplified as
	\begin{align}
		\boldsymbol{F} = \frac{2} {\sigma^2_w}  \mathcal{R}\Bigg\{ \begin{bmatrix} \boldsymbol{\Lambda}^{\text{H}} \boldsymbol{\Lambda} ,  & \boldsymbol{\Lambda}^{\text{H}} \boldsymbol{\Upsilon}  \\
			\boldsymbol{\Upsilon}^{\text{H}} \boldsymbol{\Lambda} , & \boldsymbol{\Upsilon}^{\text{H}} \boldsymbol{\Upsilon}
		\end{bmatrix}\Bigg\}.
	\end{align}

	Finally, for an unbiased estimator, the  CRB is obtained from the Fisher information matrix. For the position perturbation $\tilde{d}_n$, the CRB is given by
	\begin{align}
		\text{var}\{\tilde{d}_n\}\geq [\boldsymbol{F}^{-1}]_{n,n},
	\end{align}
	where $[\boldsymbol{F}^{-1}]_{n,n}$ denotes the $n$-th diagonal entry of the matrix $\boldsymbol{F}^{-1}$.
	The CRB of the $k$-th DOA estimation is
	\begin{align}
		\text{var}\{\theta_k\}\geq [\boldsymbol{F}^{-1}]_{N+k,N+k}.
	\end{align}

	\section{Simulation Results}\label{sec5}
	In this section, simulation results are given  to show the performance of the proposed ADPP method for the DOA estimation in the UAV swarm system with RIS, where the position perturbation is considered.  Simulation results are carried out in a personal computer with MATLAB R2020b, Intel Core i5 @ 2.9 GHz processor, and 8 GB LPDDR3 @ 2133 MHz. The MATLAB code about the ADPP method is available online \url{https://github.com/chenpengseu/ADPP.git}. Simulation parameters are given in Table~\ref{table1}. In the proposed ADPP method, the estimated position perturbation is set as $\hat{\boldsymbol{d}} = \boldsymbol{0}$ at the initialization step. Since the mean of the position perturbation caused by the UAV movement is zero, it is reasonable to set the initial value as $\boldsymbol{0}$. As given in Table~\ref{table1}, we use $32$ UAVs to form a group and $32$ measurements for the DOA estimation. The distance between adjacent UAVs is $\lambda/2$, ensuring that there is no grating lobe using the UAV array.

\begin{table}% [!t] 
	\renewcommand{\arraystretch}{1.3}
	\caption{Simulation parameters}
	\label{table1}
	\centering
	\begin{tabular}{cc}
		\hline
		\textbf{Parameter}                 & \textbf{Value}                                                   \\
		\hline
		The number of UAVs                 & $N=32$                                                           \\
		The number of measurements         & $M=32$                                                           \\
		The distance between adjacent UAVs & $d=\lambda/2$                                                    \\
		The number of targets              & $K=3$                                                            \\
		The detection range                & $[-\ang{45}, \ang{45}]$                                          \\
		The directions of targets          & $\boldsymbol{\theta} =[-\ang{30.345},\ang{0.789}, \ang{20.456}]$ \\
		\hline
	\end{tabular}
\end{table}
%The simulation parameters are shown in Table

\begin{figure}
	\centering
	\includegraphics[width=3.8in]{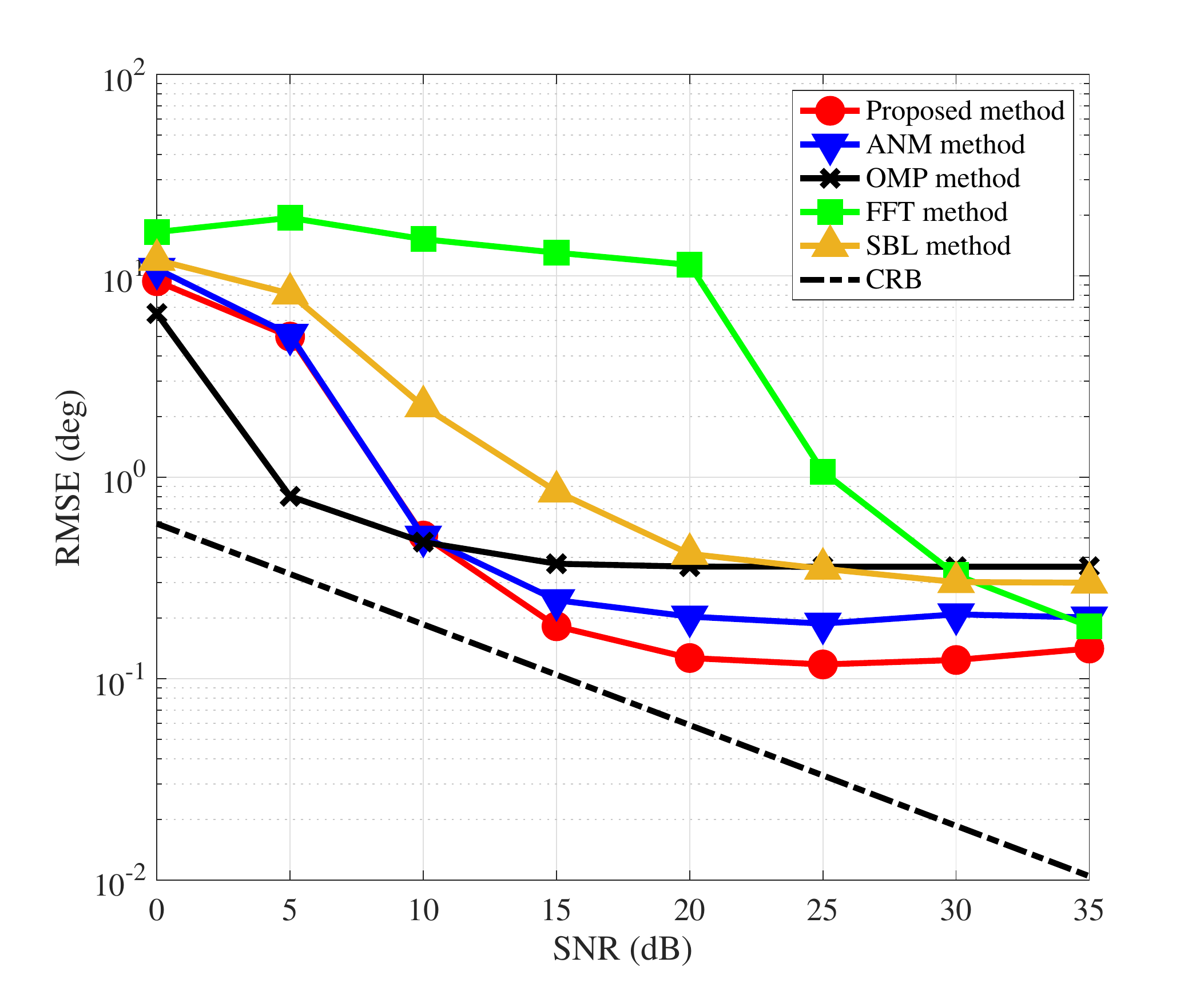}
	\caption{ The DOA estimation performance with different SNRs.}
	\label{SNR}
\end{figure}

\begin{figure}
	\centering
	\includegraphics[width=3.8in]{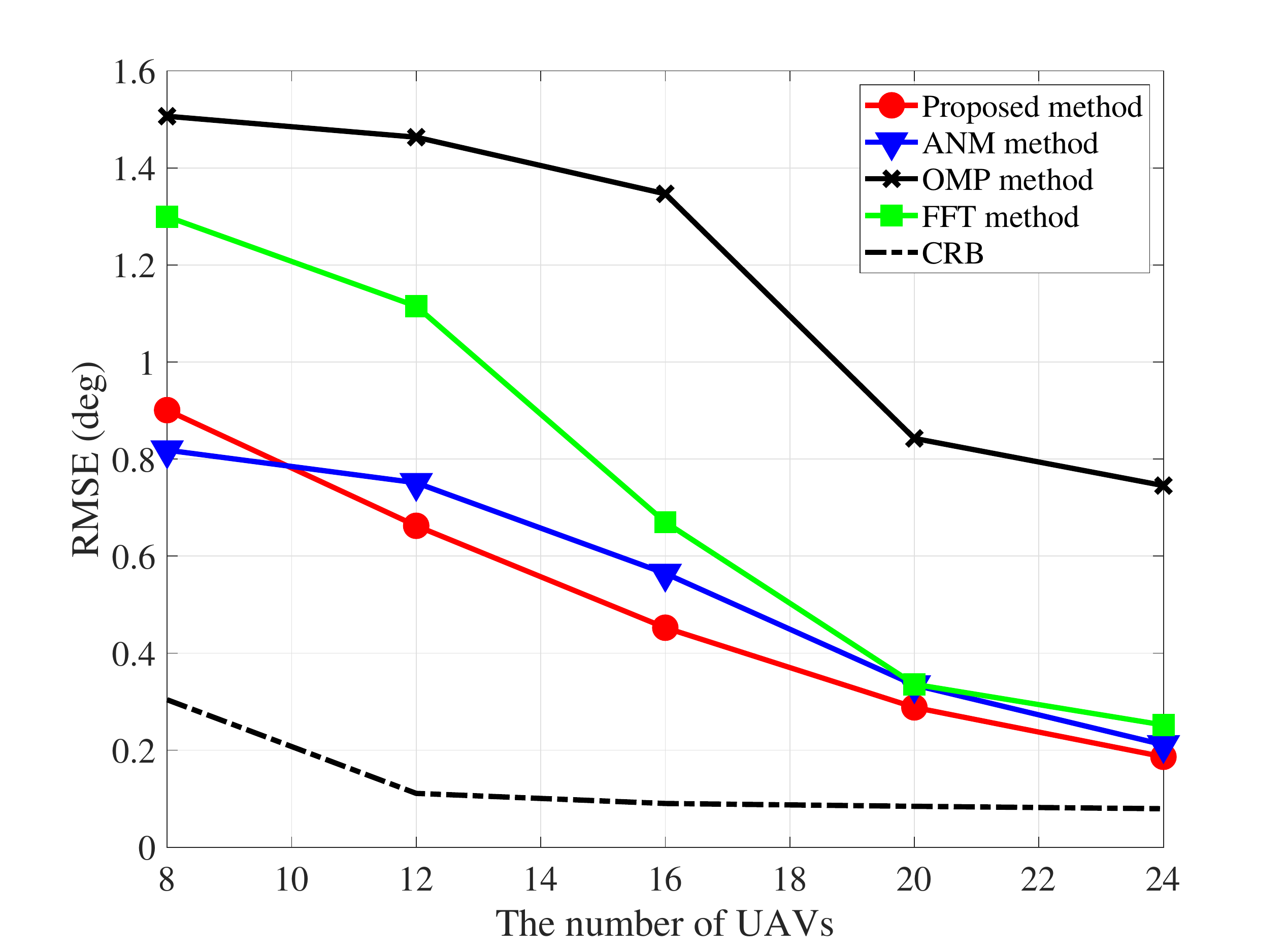}
	\caption{The DOA estimation performance with different numbers of UAVs.}
	\label{ANT}
\end{figure}

First, we show the DOA estimation performance with different SNRs, where the estimation performance is measured by the RMSE defined as
\begin{align}
\text{RMSE} = \sqrt{\frac{1}{N_{\text{MC}}K}\|\boldsymbol{\theta}-\hat{\boldsymbol{\theta}}\|^2_2},
\end{align}
where $N_{\text{MC}}$ denotes the number of simulations. The RMSE is measured in degree, and has been used in many literatures~\cite{9016105,9384289,9367250}. The RMSE measures the average error of the DOA estimation for multiple received signals. The proposed method is compared with $3$ benchmark  methods:
\begin{itemize}
	\item \emph{FFT method}: The fast Fourier transformation (FFT) method is used for the DOA estimation, and signal directions are estimated by finding peak positions.
	\item \emph{ANM method}~\cite{6576276}: The atomic norm minimization (ANM) method without the off-grid effect is used for the DOA estimation, and  the direction is estimated by the ANM-based reconstruction method.
	\item \emph{OMP method}~\cite{4385788,1337101}: The OMP is used for the DOA estimation by discretizing the spatial angle into grids, and the direction is estimated by the OMP-based reconstruction method.
	\item \emph{SBL method}~\cite{9521821,6320676}: The sparse Bayesian learning (SBL)-based method for the DOA estimation is proposed to obtain a stationary DOA estimation iteratively with the distribution assumption of the received signal.
\end{itemize}
As shown in Fig.~\ref{SNR}, the proposed method has the best estimation performance with the SNR being greater than $15$~dB and can approach the CRB. However, when the SNR is less than $15$~dB, the estimation performance of the proposed method is almost the same as the ANM-based method.

\begin{figure}
	\centering
	\includegraphics[width=3.8in]{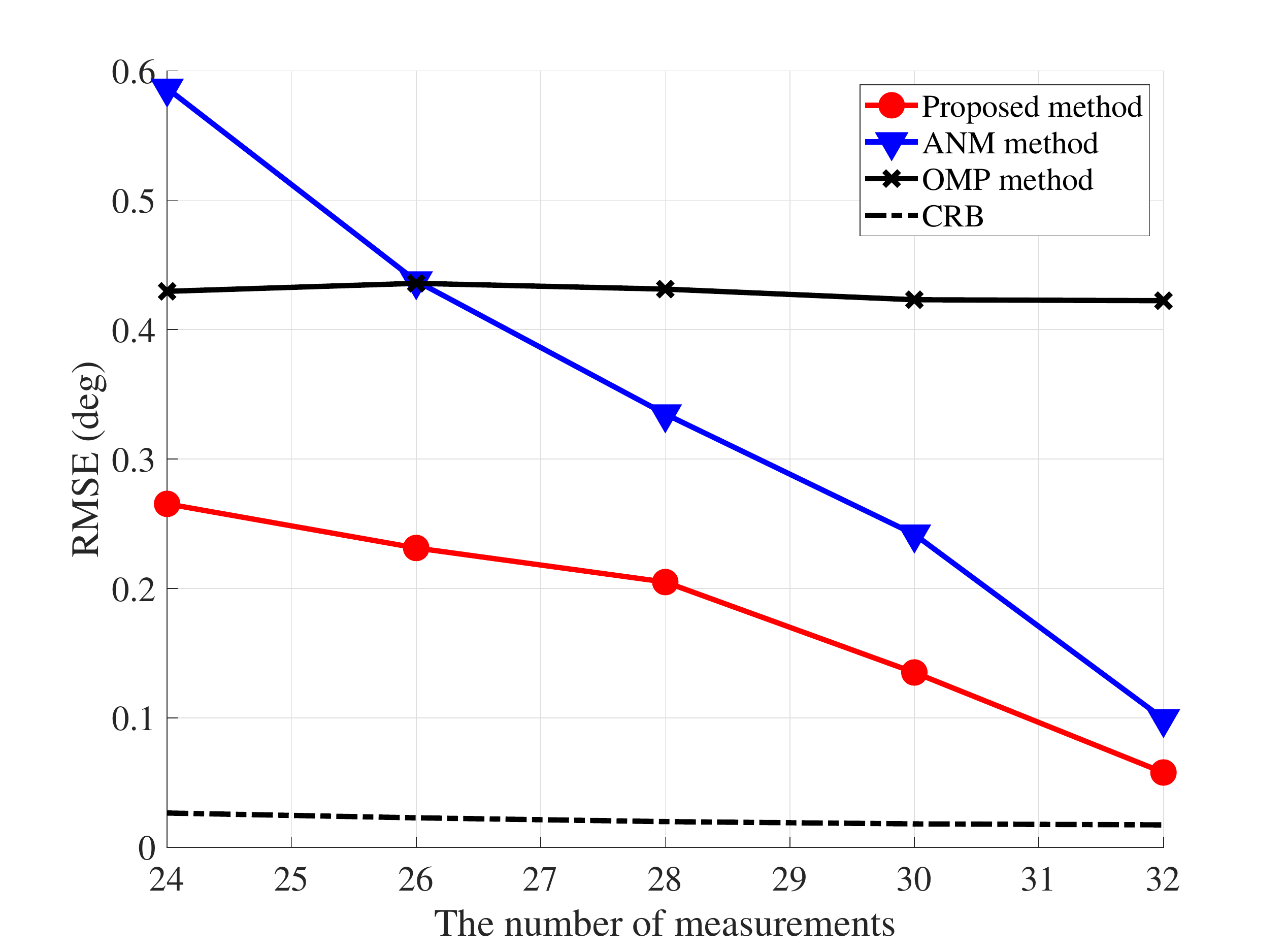}
	\caption{The DOA estimation performance with different numbers of measurements.}
	\label{MEA}
\end{figure}

\begin{figure}
	\centering
	\includegraphics[width=3.8in]{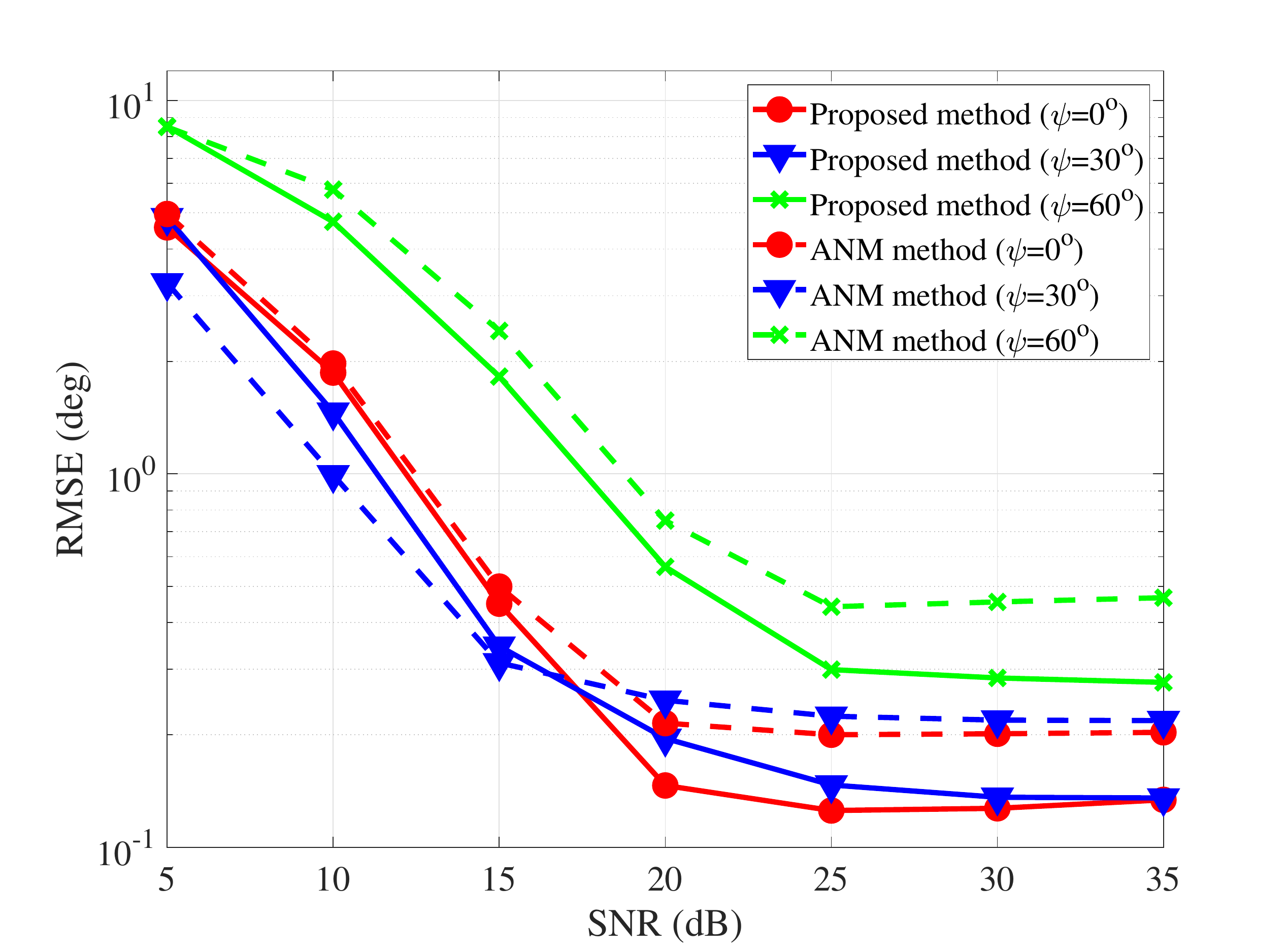}
	\caption{The DOA estimation performance with different received angles $\psi$.}
	\label{PSI}
\end{figure}

\begin{table}[!t] 
	\renewcommand{\arraystretch}{1.3}
	\caption{ Computational Time}
	\label{table2}
	\centering
	\begin{tabular}{cc}
		\hline
		\textbf{ Method} &  \textbf{ Time (s)} \\
		\hline
		 FFT method  &  $1.872$    \\
		 OMP method & $0.148$   \\
		 ANM method & $5.038$                                                  \\
		 SBL method & $13.823$\\
		 Proposed method  & $8.705$\\
		\hline
	\end{tabular}
\end{table}

Second, the proposed method is also compared with existing methods in the scenario under different numbers of UAVs, and simulation results are shown in Fig.~\ref{ANT}. The proposed method has better DOA estimation performance when the number of UAVs is no less than $12$ and can approach the CRB with more UAVs. Then, the DOA estimation performance of the proposed method concerning different numbers of measurements is shown in Fig.~\ref{MEA}, where the estimation performance approaches the CRB with more measurements. The proposed method achieves a better estimation performance than the traditional ANM method, especially in a scenario with fewer measurements.  To show the computational complexity clearly, the computational time is given in Table~\ref{table2}. The computational time of the proposed method is longer than FFT, OMP, and ANM methods but shorter than the SBL method since both the sparse reconstruction and the perturbation estimation are processed in the proposed method. Hence, for practical applications, the proposed method can be improved with lower computational complexity.

Finally, the received angle is also considered, where the receiving system gets the reflected signal from different angles. As shown in  Fig.~\ref{PSI},  the estimation performance is sensitive to the incident angle, and the better estimation performance is achieved by setting the received angle to $\ang{0}$. Furthermore, the proposed method can achieve much better estimation performance than the traditional ANM counterpart regardless of different angles.   For the limitations of the proposed method, the proposed method is based on the assumption of narrowband signal, so the proposed method cannot be used for wideband signals directly. Another limitation is the computational complexity of the proposed method, so the proposed method cannot be used for the DOA estimation of a fast-moving target.

\section{Conclusions}\label{sec6}
The DOA estimation problem has been considered in the scenario with the UAV swarm system using the RIS, where only one receiving channel was used to reduce the system cost. Then, a novel atomic norm-based estimation method has been proposed with the position perturbations of UAVs by exploiting the target sparsity in the spatial domain. By introducing the transforming matrix, the DOA estimation problem is solved by the SDP method, and the estimation results have been further refined by the gradient descent method. Simulation results have shown the performance improvement in the scenario with the position perturbation taken into account. In the future, it is an exciting direction to consider the DOA estimation method with lower computational complexity. Additionally, we will also focus on the theoretical analysis for the proposed method's convergence and the selection of hyperparameters.

\begin{appendices}
	\section{ Proof of Proposition~\ref{pr1}}\label{apd1}
	To solve the optimization problem (\ref{eq23}),  we can rewrite it as the following optimization problem
	\begin{align}
		\min_{\boldsymbol{x},\boldsymbol{z}} & \quad \frac{1}{2}\|\boldsymbol{r}-\boldsymbol{B}^{\text{T}}\text{diag}\{\boldsymbol{a}(\psi, \tilde{\boldsymbol{d}})\}\boldsymbol{x}\|^2_2+\beta \|\boldsymbol{z}\|_{\mathcal{A}} \\
		\text{s.t.}                          & \quad \boldsymbol{z}=\boldsymbol{x}.\notag
	\end{align}
	Then, by introducing a Lagrangian parameter $\boldsymbol{h}$, the corresponding Lagrangian function can be expressed as
	\begin{align}
\mathcal{L}(\boldsymbol{x},\boldsymbol{z},\boldsymbol{h}) & \triangleq \frac{1}{2}\|\boldsymbol{r}-\boldsymbol{B}^{\text{T}}\text{diag}\{\boldsymbol{a}(\psi, \tilde{\boldsymbol{d}})\}\boldsymbol{x}\|^2_2+\beta \|\boldsymbol{z}\|_{\mathcal{A}}\notag \\
& \qquad +\langle \boldsymbol{h},\boldsymbol{x}-\boldsymbol{z} \rangle,
\end{align}
	where $\langle \boldsymbol{h},\boldsymbol{x}-\boldsymbol{z} \rangle$ is defined as
	\begin{align}
		\langle \boldsymbol{h},\boldsymbol{x}-\boldsymbol{z} \rangle\triangleq \mathcal{R}\{(\boldsymbol{x}-\boldsymbol{z})^{\text{H}}\boldsymbol{h}\}.
	\end{align}
	Hence, the dual function can be obtained as
	\begin{align}
		g(\boldsymbol{h}) & \triangleq \inf_{\boldsymbol{x},\boldsymbol{z}} \mathcal{L}(\boldsymbol{x},\boldsymbol{z},\boldsymbol{h})\label{eq15}                                                                                                               \\
		                  & = \inf_{\boldsymbol{x}}\left(\frac{1}{2}\|\boldsymbol{r}-\boldsymbol{B}^{\text{T}}\text{diag}\{\boldsymbol{a}(\psi, \tilde{\boldsymbol{d}})\}\boldsymbol{x}\|^2_2+\langle \boldsymbol{h},\boldsymbol{x}\rangle\right)\notag         \\
		                  & \qquad +\inf_{\boldsymbol{z}} \left(\beta \|\boldsymbol{z}\|_{\mathcal{A}}-\langle \boldsymbol{h},\boldsymbol{z} \rangle\right)\notag                                                                                               \\
		                  & = \frac{1}{2}\bigg[\|\boldsymbol{r}\|^2_2-(\text{diag}\{\boldsymbol{a}^{*}(\psi, \tilde{\boldsymbol{d}})\}\boldsymbol{B}^{\text{*}}\boldsymbol{r}-\boldsymbol{h})^{\text{H}}\notag                                                  \\
		                  & \qquad (\text{diag}\{\boldsymbol{a}^{*}(\psi, \tilde{\boldsymbol{d}})\}\boldsymbol{B}^{\text{*}}\boldsymbol{B}^{\text{T}}\text{diag}\{\boldsymbol{a}(\psi, \tilde{\boldsymbol{d}})\})^{-1}\notag                                    \\
		                  & \qquad(\text{diag}\{\boldsymbol{a}^{*}(\psi, \tilde{\boldsymbol{d}})\}\boldsymbol{B}^{\text{*}}\boldsymbol{r}-\boldsymbol{h})\bigg]-I_{\{\boldsymbol{h}:\|\boldsymbol{h}\|_{\tilde{\mathcal{A}}}\leq\beta\}}(\boldsymbol{h}),\notag
	\end{align}
	where an indicator function is defined as
	\begin{align}
		I_{\mathcal{S}}(x)=\begin{cases}
			0, & x\in\mathcal{S}  \\
			\infty, & \text{otherwise}
		\end{cases}.
	\end{align}
	Moreover, in (\ref{eq15}), we define a dual norm of the atomic norm as
	\begin{align}
		\|\boldsymbol{h}\|_{\tilde{A}}\triangleq \sup_{\|\boldsymbol{u}\|_{\mathcal{A}}\leq 1}\langle \boldsymbol{u}, \boldsymbol{h}
		\rangle.
	\end{align}
	Then, the optimization problem (\ref{eq23}) can be converted into a dual problem
	\begin{align}
		\max_{\boldsymbol{h}} & \quad \frac{1}{2}\bigg[\|\boldsymbol{r}\|^2_2-(\text{diag}\{\boldsymbol{a}^{*}(\psi, \tilde{\boldsymbol{d}})\}\boldsymbol{B}^{\text{*}}\boldsymbol{r}-\boldsymbol{h})^{\text{H}}\notag           \\
& \qquad (\text{diag}\{\boldsymbol{a}^{*}(\psi, \tilde{\boldsymbol{d}})\}\boldsymbol{B}^{\text{*}}\boldsymbol{B}^{\text{T}}\text{diag}\{\boldsymbol{a}(\psi, \tilde{\boldsymbol{d}})\})^{-1}\notag \\
& \qquad(\text{diag}\{\boldsymbol{a}^{*}(\psi, \tilde{\boldsymbol{d}})\}\boldsymbol{B}^{\text{*}}\boldsymbol{r}-\boldsymbol{h})\bigg]\label{eq19}                                                  \\
		\text{s.t.}           & \quad \|\boldsymbol{h}\|_{\tilde{\mathcal{A}}}\leq \beta,\notag
	\end{align}
	which is equal to the following problem
	\begin{align}
		\min_{\boldsymbol{h}} & \quad \left[\text{diag}\{\boldsymbol{a}^{*}(\psi, \tilde{\boldsymbol{d}})\}\boldsymbol{B}^{\text{*}}\boldsymbol{r}-\boldsymbol{h}\right]^{\text{H}}\notag                                                   \\
& \qquad \left[\text{diag}\{\boldsymbol{a}^{*}(\psi, \tilde{\boldsymbol{d}})\}\boldsymbol{B}^{\text{*}}\boldsymbol{B}^{\text{T}}\text{diag}\{\boldsymbol{a}(\psi, \tilde{\boldsymbol{d}})\}\right]^{-1}\notag \\
& \qquad\left[\text{diag}\{\boldsymbol{a}^{*}(\psi, \tilde{\boldsymbol{d}})\}\boldsymbol{B}^{\text{*}}\boldsymbol{r}-\boldsymbol{h}\right]\label{eq29}                                                        \\
\text{s.t.}           & \quad \|\boldsymbol{h}\|_{\tilde{\mathcal{A}}}\leq \beta.\notag
	\end{align}

	For the dual norm $\|\boldsymbol{h}\|_{\tilde{A}}$, we can simplified it as
\begin{align}
 & \|\boldsymbol{h}\|_{\tilde{A}} =\sup_{\|\boldsymbol{u}\|_{\mathcal{A}}\leq 1}\langle \boldsymbol{u}, \boldsymbol{h}
\rangle\notag                                                                                                                 \\
 & \quad = \sup_{\substack{
		c_i>0, \phi_i\in[0,2\pi)                                                                                                      \\\theta_i\in\left(-\frac{\pi}{2},\frac{\pi}{2}\right],
		\sum_i c_i\leq 1}
}
\left \langle
\sum_i c_i e^{j\phi_i}\boldsymbol{a}(\theta_i,\tilde{\boldsymbol{d}})\odot \boldsymbol{a}(\theta_i,\bar{\boldsymbol{d}}),
\boldsymbol{h}
\right \rangle\notag                                                                                                          \\
 & \quad = \sup_{\substack{
		c_i>0, \phi_i\in[0,2\pi)                                                                                                      \\\theta_i\in\left(-\frac{\pi}{2},\frac{\pi}{2}\right],
		\sum_i c_i\leq 1}
}
\left \langle
\sum_i c_i e^{j\phi_i} \boldsymbol{a}(\theta_i,\tilde{\boldsymbol{d}}+\bar{\boldsymbol{d}}),
\boldsymbol{h}
\right \rangle\notag                                                                                                          \\
 & \quad = \sup_{\substack{
		c_i>0, \phi_i\in[0,2\pi)                                                                                                      \\\theta_i\in\left(-\frac{\pi}{2},\frac{\pi}{2}\right],
		\sum_i c_i\leq 1}
}\sum_i \mathcal{R}\left\{
\boldsymbol{h}^{\text{H}} c_i e^{j\phi_i} \boldsymbol{a}(\theta_i,\tilde{\boldsymbol{d}}+\bar{\boldsymbol{d}})\right\} \notag \\
 & \quad = \sup_{\substack{
		c_i>0, \theta_i\in\left(-\frac{\pi}{2},\frac{\pi}{2}\right]                                                                   \\
		\sum_i c_i\leq 1}
}\sum_i c_i \left|
\boldsymbol{x}^{\text{H}} \boldsymbol{a}(\theta_i,\tilde{\boldsymbol{d}}+\bar{\boldsymbol{d}})\right| \notag                  \\
 & \quad = \sup_{
	\theta\in\left(-\frac{\pi}{2},\frac{\pi}{2}\right]	} \left|
\boldsymbol{h}^{\text{H}} \boldsymbol{a}(\theta,\tilde{\boldsymbol{d}}+\bar{\boldsymbol{d}})\right|.\label{eq31}
\end{align}
Then, the constraint in (\ref{eq29}) can be rewritten as
\begin{align}
	\sup_{
		\theta\in\left(-\frac{\pi}{2},\frac{\pi}{2}\right]	} \left|
	\boldsymbol{h}^{\text{H}} \boldsymbol{a}(\theta,\tilde{\boldsymbol{d}}+\bar{\boldsymbol{d}})\right|\leq \beta.
\end{align}
To further simplify the constraint, we consider to use a transformation matrix $\boldsymbol{T}\in\mathbb{C}^{N\times N}$ to convert the vector  $\boldsymbol{a}(\theta,\bar{\boldsymbol{d}})$ into $\boldsymbol{a}(\theta,\tilde{\boldsymbol{d}}+\bar{\boldsymbol{d}})$, i.e.,
\begin{align}
	\boldsymbol{T}^{\text{H}}\boldsymbol{a}(\theta,\bar{\boldsymbol{d}})=\boldsymbol{a}(\theta,\tilde{\boldsymbol{d}}+\bar{\boldsymbol{d}}).
\end{align}
Then, we can build a Hermitian matrix $
	\begin{bmatrix}
		\boldsymbol{W}               & \boldsymbol{Th} \\
		(\boldsymbol{Th})^{\text{H}} & t
	\end{bmatrix}$,
where $t\in\mathbb{R}$ is hyperparameter and $\boldsymbol{W}$ is a Hermitian matrix. This matrix is a semi-definite positive matrix, i.e.,
\begin{align}
	\begin{bmatrix}
		\boldsymbol{W}               & \boldsymbol{Th} \\
		(\boldsymbol{Th})^{\text{H}} & t
	\end{bmatrix}\succeq 0,
\end{align}
if and only if we have
\begin{align}
	 & \boldsymbol{W}\succeq 0,                                                                                      \\
	 & \boldsymbol{W}-t^{-1} \boldsymbol{Th}\boldsymbol{h}^{\text{H}}\boldsymbol{T}^{\text{H}}\succeq 0.\label{eq26}
\end{align}
Hence, for any given vector $\boldsymbol{b}\in\mathbb{C}^{N\times 1}$ with (\ref{eq26}), we have
\begin{align}
	\boldsymbol{b}^{\text{H}}\left(\boldsymbol{W}-t^{-1} \boldsymbol{Th}\boldsymbol{h}^{\text{H}}\boldsymbol{T}^{\text{H}}\right)\boldsymbol{b}\geq 0,
\end{align}
which shows that
\begin{align}
	|\boldsymbol{h}^{\text{H}}\boldsymbol{T}^{\text{H}}\boldsymbol{b}|^2\leq t \boldsymbol{b}^{\text{H}}\boldsymbol{W}\boldsymbol{b}.\label{eq30}
\end{align}

By choosing $\boldsymbol{b}=\boldsymbol{a}(\theta, \bar{\boldsymbol{d}})$, we can rewrite (\ref{eq30}) as
\begin{align}
	 & \left|
	\boldsymbol{h}^{\text{H}} \boldsymbol{T}^{\text{H}}\boldsymbol{a}(\theta, \bar{\boldsymbol{d}})\right|^2 \leq t \boldsymbol{a}^{\text{H}}(\theta, \bar{\boldsymbol{d}})\boldsymbol{W}\boldsymbol{a}(\theta, \bar{\boldsymbol{d}})\label{eq37} \\
	 & \qquad = t\sum_{n_1}\sum_{n_2}a^{\text{*}}(\theta, \bar{d}_{n_1})W_{n_1,n_2}a(\theta, \bar{d}_{n_2})\notag                                                                                                                                 \\
	 & \qquad = t\text{Tr}\{ \boldsymbol{W}\}+t\sum_{\nu\neq 0} e^{j2\nu\pi/\lambda \sin\theta}\sum_{n}
	W_{n,n+\nu}. \notag
\end{align}
For the matrix $\boldsymbol{W}$, if we have
\begin{align}
	 & \text{Tr}\{\boldsymbol{W}\} = \beta^2/t, \\
	 & \sum_{n} W_{n,n+\nu} = 0, \nu\neq 0,
\end{align}
with the transformation matrix $\boldsymbol{T}$, (\ref{eq37}) indicates that
\begin{align}
	\left|
	\boldsymbol{h}^{\text{H}} \boldsymbol{T}^{\text{H}}\boldsymbol{a}(\theta, \bar{\boldsymbol{d}})\right|^2=\left|
	\boldsymbol{h}^{\text{H}} \boldsymbol{a}(\theta,\tilde{\boldsymbol{d}}+\bar{\boldsymbol{d}})\right|^2 \leq \beta^2.\label{eq32}
\end{align}
Combining with (\ref{eq31}), the dual norm condition $\|\boldsymbol{h}\|_{\tilde{A}}\leq \beta$ can be satisfied.

Finally, the constraint of the optimization problem (\ref{eq29}) can be expressed as a SDP constraint
\begin{align}
	 & \begin{bmatrix}
		\boldsymbol{W}               & \boldsymbol{Th} \\
		(\boldsymbol{Th})^{\text{H}} & t
	\end{bmatrix}\succeq 0,\notag     \\
	 & \boldsymbol{W} \text{ is Hermitian},           \\
	 & \text{Tr}\{\boldsymbol{W}\} = \beta^2/t,\notag \\
	 & \sum_{n} W_{n,n+\nu} = 0, \nu\neq 0.\notag
\end{align}
Then, the SDP in Proposition~\ref{pr1} is proved.
\end{appendices}

\bibliographystyle{IEEEtran}
\bibliography{IEEEabrv.bib,References.bib}

\begin{thebibliography}{10}
\providecommand{\url}[1]{#1}
\csname url@samestyle\endcsname
\providecommand{\newblock}{\relax}
\providecommand{\bibinfo}[2]{#2}
\providecommand{\BIBentrySTDinterwordspacing}{\spaceskip=0pt\relax}
\providecommand{\BIBentryALTinterwordstretchfactor}{4}
\providecommand{\BIBentryALTinterwordspacing}{\spaceskip=\fontdimen2\font plus
\BIBentryALTinterwordstretchfactor\fontdimen3\font minus
  \fontdimen4\font\relax}
\providecommand{\BIBforeignlanguage}[2]{{%
\expandafter\ifx\csname l@#1\endcsname\relax
\typeout{** WARNING: IEEEtran.bst: No hyphenation pattern has been}%
\typeout{** loaded for the language `#1'. Using the pattern for}%
\typeout{** the default language instead.}%
\else
\language=\csname l@#1\endcsname
\fi
#2}}
\providecommand{\BIBdecl}{\relax}
\BIBdecl

\bibitem{9541018}
Z.~Zheng, Y.~Huang, W.-Q. Wang, and H.~C. So, ``Augmented covariance matrix
  reconstruction for {DOA} estimation using difference coarray,'' \emph{{IEEE}
  Trans. Signal Process.}, vol.~69, pp. 5345--5358, 2021.

\bibitem{9384289}
M.~Wagner, Y.~Park, and P.~Gerstoft, ``Gridless {DOA} estimation and
  root-{MUSIC} for non-uniform linear arrays,'' \emph{{IEEE} Trans. Signal
  Process.}, vol.~69, pp. 2144--2157, 2021.

\bibitem{6747980}
F.~Wen, Q.~Wan, R.~Fan, and H.~Wei, ``Improved {MUSIC} algorithm for multiple
  noncoherent subarrays,'' \emph{{IEEE} Signal Process. Lett.}, vol.~21, no.~5,
  pp. 527--530, 2014.

\bibitem{1395953}
F.~Gao and A.~Gershman, ``A generalized {ESPRIT} approach to
  direction-of-arrival estimation,'' \emph{{IEEE} Signal Process. Lett.},
  vol.~12, no.~3, pp. 254--257, 2005.

\bibitem{7226785}
J.~Lin, X.~Ma, S.~Yan, and C.~Hao, ``Time-frequency multi-invariance {ESPRIT
  for DOA} estimation,'' \emph{{IEEE} Antennas Wireless Propag. Lett.},
  vol.~15, pp. 770--773, 2016.

\bibitem{9174801}
X.~Liu, Y.~Liu, Y.~Chen, and H.~V. Poor, ``{RIS} enhanced massive
  non-orthogonal multiple access networks: {Deployment} and passive beamforming
  design,'' \emph{{IEEE} J. Sel. Areas Commun.}, vol.~39, no.~4, pp.
  1057--1071, 2021.

\bibitem{9309152}
L.~You, J.~Xiong, D.~W.~K. Ng, C.~Yuen, W.~Wang, and X.~Gao, ``Energy
  efficiency and spectral efficiency tradeoff in {RIS}-aided multiuser {MIMO}
  uplink transmission,'' \emph{{IEEE} Trans. Signal Process.}, vol.~69, pp.
  1407--1421, 2021.

\bibitem{9201413}
S.~Zeng, H.~Zhang, B.~Di, Z.~Han, and L.~Song, ``Reconfigurable intelligent
  surface {(RIS)} assisted wireless coverage extension: {RIS} orientation and
  location optimization,'' \emph{{IEEE} Commun. Lett.}, vol.~25, no.~1, pp.
  269--273, 2021.

\bibitem{9144510}
L.~Yang, F.~Meng, Q.~Wu, D.~B. da~Costa, and M.-S. Alouini, ``Accurate
  closed-form approximations to channel distributions of {RIS}-aided wireless
  systems,'' \emph{{IEEE Wireless Commun. Lett.}}, vol.~9, no.~11, pp.
  1985--1989, 2020.

\bibitem{9124704}
L.~Yang, F.~Meng, J.~Zhang, M.~O. Hasna, and M.~D. Renzo, ``On the performance
  of {RIS}-assisted dual-hop {UAV} communication systems,'' \emph{{IEEE} Trans.
  Veh. Technol.}, vol.~69, no.~9, pp. 10\,385--10\,390, 2020.

\bibitem{9354904}
K.~Ardah, S.~Gherekhloo, A.~L.~F. de~Almeida, and M.~Haardt, ``{TRICE}: {A}
  channel estimation framework for {RIS}-aided millimeter-wave {MIMO}
  systems,'' \emph{{IEEE} Signal Process. Lett.}, vol.~28, pp. 513--517, 2021.

\bibitem{9359337}
S.~Uemura, K.~Nishimori, R.~Taniguchi, M.~Inomata, K.~Kitao, T.~Imai,
  S.~Suyama, H.~Ishikawa, and Y.~Oda, ``{Direction-of-arrival estimation with
  circular array using compressed sensing in 20 GHz band},'' \emph{{IEEE}
  Antennas Wireless Propag. Lett.}, vol.~20, no.~5, pp. 703--707, 2021.

\bibitem{9296231}
M.~Ferreira Da~Costa and Y.~Chi, ``Compressed super-resolution of positive
  sources,'' \emph{{IEEE} Signal Process. Lett.}, vol.~28, pp. 56--60, 2021.

\bibitem{4385788}
J.~A. Tropp and A.~C. Gilbert, ``Signal recovery from random measurements via
  orthogonal matching pursuit,'' \emph{{IEEE} Trans. Inf. Theory}, vol.~53,
  no.~12, pp. 4655--4666, 2007.

\bibitem{1337101}
J.~Tropp, ``Greed is good: {Algorithmic} results for sparse approximation,''
  \emph{{IEEE} Trans. Inf. Theory}, vol.~50, no.~10, pp. 2231--2242, 2004.

\bibitem{1459044}
J.~Yedidia, W.~Freeman, and Y.~Weiss, ``Constructing free-energy approximations
  and generalized belief propagation algorithms,'' \emph{{IEEE} Trans. Inf.
  Theory}, vol.~51, no.~7, pp. 2282--2312, 2005.

\bibitem{6556987}
J.~P. Vila and P.~Schniter, ``Expectation-maximization gaussian-mixture
  approximate message passing,'' \emph{{IEEE} Trans. Signal Process.}, vol.~61,
  no.~19, pp. 4658--4672, 2013.

\bibitem{9016105}
Y.~Chi and M.~Ferreira Da~Costa, ``Harnessing sparsity over the continuum:
  {Atomic} norm minimization for superresolution,'' \emph{{IEEE} Signal
  Process. Mag.}, vol.~37, no.~2, pp. 39--57, 2020.

\bibitem{7313018}
Y.~Li and Y.~Chi, ``Off-the-grid line spectrum denoising and estimation with
  multiple measurement vectors,'' \emph{{IEEE} Trans. Signal Process.},
  vol.~64, no.~5, pp. 1257--1269, 2016.

\bibitem{6576276}
G.~Tang, B.~N. Bhaskar, P.~Shah, and B.~Recht, ``Compressed sensing off the
  grid,'' \emph{{IEEE} Trans. Inf. Theory}, vol.~59, no.~11, pp. 7465--7490,
  2013.

\bibitem{7307118}
L.~Sun, H.~Hong, Y.~Li, C.~Gu, F.~Xi, C.~Li, and X.~Zhu, ``Noncontact vital
  sign detection based on stepwise atomic norm minimization,'' \emph{{IEEE}
  Signal Process. Lett.}, vol.~22, no.~12, pp. 2479--2483, 2015.

\bibitem{7091875}
S.~Pejoski and V.~Kafedziski, ``Estimation of sparse time dispersive channels
  in pilot aided {OFDM} using atomic norm,'' \emph{{IEEE Antennas Wireless
  Propag. Lett.}}, vol.~4, no.~4, pp. 397--400, 2015.

\bibitem{8432470}
Y.~Tsai, L.~Zheng, and X.~Wang, ``Millimeter-wave beamformed full-dimensional
  {MIMO} channel estimation based on atomic norm minimization,'' \emph{{IEEE}
  Trans. Commun.}, vol.~66, no.~12, pp. 6150--6163, 2018.

\bibitem{7917313}
O.~Teke and P.~P. Vaidyanathan, ``On the role of the bounded lemma in the {SDP}
  formulation of atomic norm problems,'' \emph{{IEEE} Signal Process. Lett.},
  vol.~24, no.~7, pp. 972--976, 2017.

\bibitem{9146196}
P.~Chen, Z.~Chen, Z.~Cao, and X.~Wang, ``A new atomic norm for {DOA} estimation
  with gain-phase errors,'' \emph{{IEEE} Trans. Signal Process.}, vol.~68, pp.
  4293--4306, 2020.

\bibitem{7314978}
Z.~Yang and L.~Xie, ``Enhancing sparsity and resolution via reweighted atomic
  norm minimization,'' \emph{{IEEE} Trans. Signal Process.}, vol.~64, no.~4,
  pp. 995--1006, 2016.

\bibitem{6998075}
Y.~Chi and Y.~Chen, ``Compressive two-dimensional harmonic retrieval via atomic
  norm minimization,'' \emph{{IEEE} Trans. Signal Process.}, vol.~63, no.~4,
  pp. 1030--1042, 2015.

\bibitem{9447907}
X.~Zhang, H.~Wang, V.~Stojanovic, P.~Cheng, S.~He, X.~Luan, and F.~Liu,
  ``{Asynchronous fault detection for interval type-2 fuzzy nonhomogeneous
  higher-level Markov jump systems with uncertain transition probabilities},''
  \emph{{IEEE} Trans. Fuzzy Syst.}, pp. 1--1, 2021.

\bibitem{iet-cta}
H.~Tao, J.~Li, Y.~Chen, V.~Stojanovic, and H.~Yang, ``Robust point-to-point
  iterative learning control with trial-varying initial conditions,''
  \emph{{IET Control Theory \& Applications}}, vol.~14, no.~19, pp. 3344--3350,
  2020.

\bibitem{rnc}
H.~Fang, G.~Zhu, V.~Stojanovic, R.~Nie, S.~He, X.~Luan, and F.~Liu, ``{Adaptive
  optimization algorithm for nonlinear Markov jump systems with partial unknown
  dynamics},'' \emph{{International Journal of Robust and Nonlinear Control}},
  vol.~31, no.~6, pp. 2126--2140, 2021.

\bibitem{CHENG2021107353}
P.~Cheng, M.~Chen, V.~Stojanovic, and S.~He, ``{Asynchronous fault detection
  filtering for piecewise homogenous Markov jump linear systems via a dual
  hidden Markov model},'' \emph{{Mechanical Systems and Signal Processing}},
  vol. 151, p. 107353, 2021.

\bibitem{5456168}
P.~Pal and P.~P. Vaidyanathan, ``Nested arrays: {A} novel approach to array
  processing with enhanced degrees of freedom,'' \emph{{IEEE} Trans. Signal
  Process.}, vol.~58, no.~8, pp. 4167--4181, 2010.

\bibitem{9411879}
S.~Liu, Z.~Mao, Y.~D. Zhang, and Y.~Huang, ``Rank minimization-based toeplitz
  reconstruction for {DoA} estimation using coprime array,'' \emph{{IEEE}
  Commun. Lett.}, vol.~25, no.~7, pp. 2265--2269, 2021.

\bibitem{9367250}
M.~Fu, Z.~Zheng, W.-Q. Wang, and H.~C. So, ``Coarray interpolation for {DOA}
  estimation using coprime {EMVS} array,'' \emph{{IEEE} Signal Process. Lett.},
  vol.~28, pp. 548--552, 2021.

\bibitem{9442317}
S.~Li and X.-P. Zhang, ``Dilated arrays: {A} family of sparse arrays with
  increased uniform degrees of freedom and reduced mutual coupling on a moving
  platform,'' \emph{{IEEE} Trans. Signal Process.}, vol.~69, pp. 3367--3382,
  2021.

\bibitem{9432742}
A.~Barthelme and W.~Utschick, ``A machine learning approach to {DoA} estimation
  and model order selection for antenna arrays with subarray sampling,''
  \emph{{IEEE} Trans. Signal Process.}, vol.~69, pp. 3075--3087, 2021.

\bibitem{9400719}
------, ``{DoA} estimation using neural network-based covariance matrix
  reconstruction,'' \emph{{IEEE} Signal Process. Lett.}, vol.~28, pp. 783--787,
  2021.

\bibitem{9442863}
G.~Yao, H.~Zhang, L.~Li, and F.~Hu, ``The {ORLS}-based {DoA} estimation for
  unknown mixtures of uncorrelated and coherent signals under unknown number of
  sources,'' \emph{{IEEE} Signal Process. Lett.}, vol.~28, pp. 1105--1109,
  2021.

\bibitem{9410582}
Y.~Mao, G.~Zhang, and H.~Leung, ``Harmonic retrieval joint multiple regression:
  {Robust DOA} estimation for {FMCW} radar in the presence of unknown spatially
  colored noise,'' \emph{{IEEE} Commun. Lett.}, vol.~25, no.~7, pp. 2240--2244,
  2021.

\bibitem{cvx}
M.~Grant and S.~Boyd, ``{CVX}: {Matlab} software for disciplined convex
  programming, version 2.1,'' Mar. 2014.

\bibitem{gb08}
------, ``Graph implementations for nonsmooth convex programs,'' in
  \emph{Recent Advances in Learning and Control}, ser. Lecture Notes in Control
  and Information Sciences, V.~Blondel, S.~Boyd, and H.~Kimura, Eds.\hskip 1em
  plus 0.5em minus 0.4em\relax Springer-Verlag Limited, 2008, pp. 95--110.

\bibitem{9410615}
L.~Shi and Y.~Chi, ``Manifold gradient descent solves multi-channel sparse
  blind deconvolution provably and efficiently,'' \emph{{IEEE} Trans. Inf.
  Theory}, vol.~67, no.~7, pp. 4784--4811, 2021.

\bibitem{9398573}
T.~Tong, C.~Ma, and Y.~Chi, ``Low-rank matrix recovery with scaled subgradient
  methods: {Fast} and robust convergence without the condition number,''
  \emph{{IEEE} Trans. Signal Process.}, vol.~69, pp. 2396--2409, 2021.

\bibitem{kalantari2020equivalence}
B.~{Kalantari}, ``{On the Equivalence of SDP Feasibility and a Convex Hull
  Relaxation for System of Quadratic Equations},'' \emph{arXiv e-prints}, p.
  arXiv:1911.03989, Nov. 2019.

\bibitem{9449977}
Y.~Liang, W.~Cui, Q.~Shen, W.~Liu, and H.~Wu, ``Cram\'{e}r-{Rao} bound for
  {DOA} estimation exploiting multiple frequency pairs,'' \emph{{IEEE} Signal
  Process. Lett.}, vol.~28, pp. 1210--1214, 2021.

\bibitem{9521821}
J.~Dai and H.~C. So, ``Real-valued sparse {Bayesian} learning for {DOA}
  estimation with arbitrary linear arrays,'' \emph{{IEEE} Trans. Signal
  Process.}, vol.~69, pp. 4977--4990, 2021.

\bibitem{6320676}
Z.~Yang, L.~Xie, and C.~Zhang, ``Off-grid direction of arrival estimation using
  sparse {Bayesian} inference,'' \emph{{IEEE} Trans. Signal Process.}, vol.~61,
  no.~1, pp. 38--43, 2013.

\end{thebibliography}

\begin{IEEEbiography}[{\includegraphics[width=1in,height=1.25in,clip,keepaspectratio]{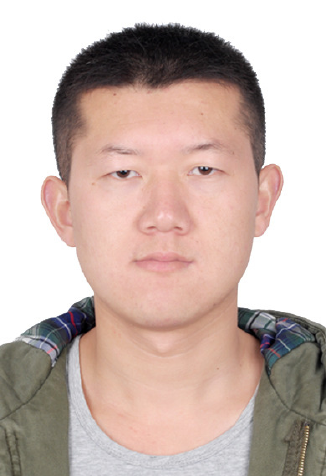}}]{Peng Chen (S'15-M'17)}
	was born in Jiangsu, China in 1989. He received the B.E. degree in 2011 and the Ph.D. degree in 2017, both from the School of Information Science and Engineering, Southeast University, China. He is currently an associate professor at the State Key Laboratory of Millimeter Waves, Southeast University. His research interests include target localization, super-resolution reconstruction, and array signal processing.
	
	From Mar. 2015 to Apr. 2016, he was a Visiting Scholar in the Electrical Engineering Department, Columbia University, New York, NY, USA.  He was a recipient of the Best Paper Award from the IEEE International Conference on Communication, Control, Computing and Electronic Engineering in 2017. 
\end{IEEEbiography}

\begin{IEEEbiography}[{\includegraphics[width=1in,height=1.25in,clip,keepaspectratio]{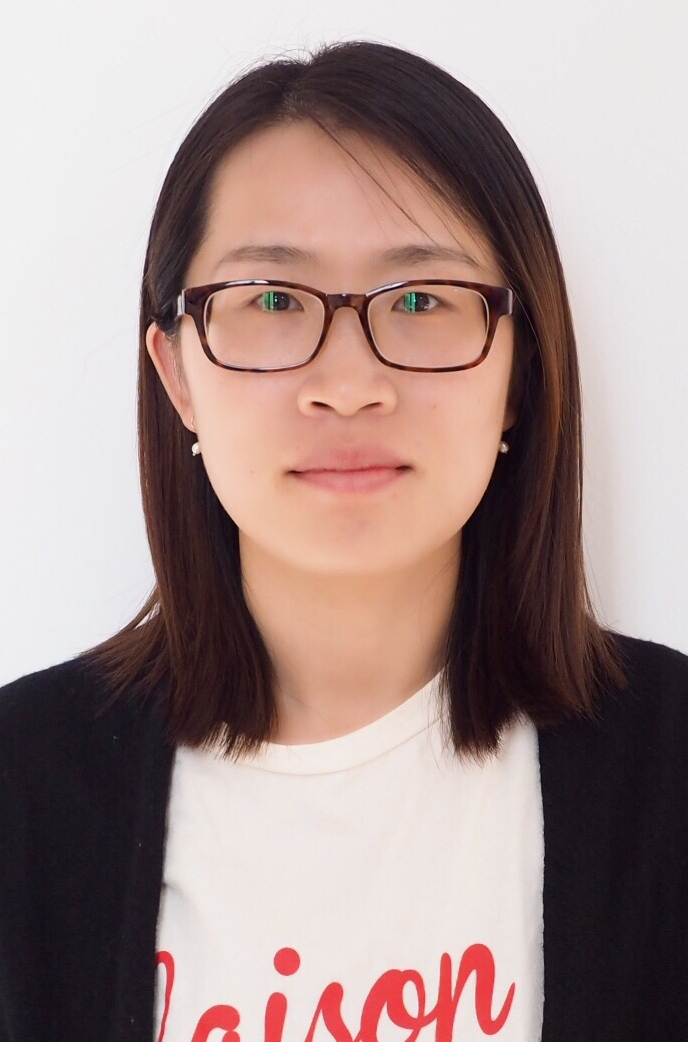}}]{Zhimin Chen (M'17)}  was born in Shandong, China, in 1985. She received the Ph.D. degree in information and communication engineering from the School of Information Science and Engineering, Southeast University, Nanjing, China in 2015. 
	
Since 2015, she is currently an associate professor at Shanghai Dianji University, Shanghai, China. From 2021,  she is also a Visiting Scholar in  the Department of Electronic and Information Engineering, The Hong Kong Polytechnic University, Hong Kong.  Her research interests include array signal processing, vehicle communications and millimeter-wave communications. 
\end{IEEEbiography}

\begin{IEEEbiography}[{\includegraphics[width=1in,height=1.25in,clip,keepaspectratio]{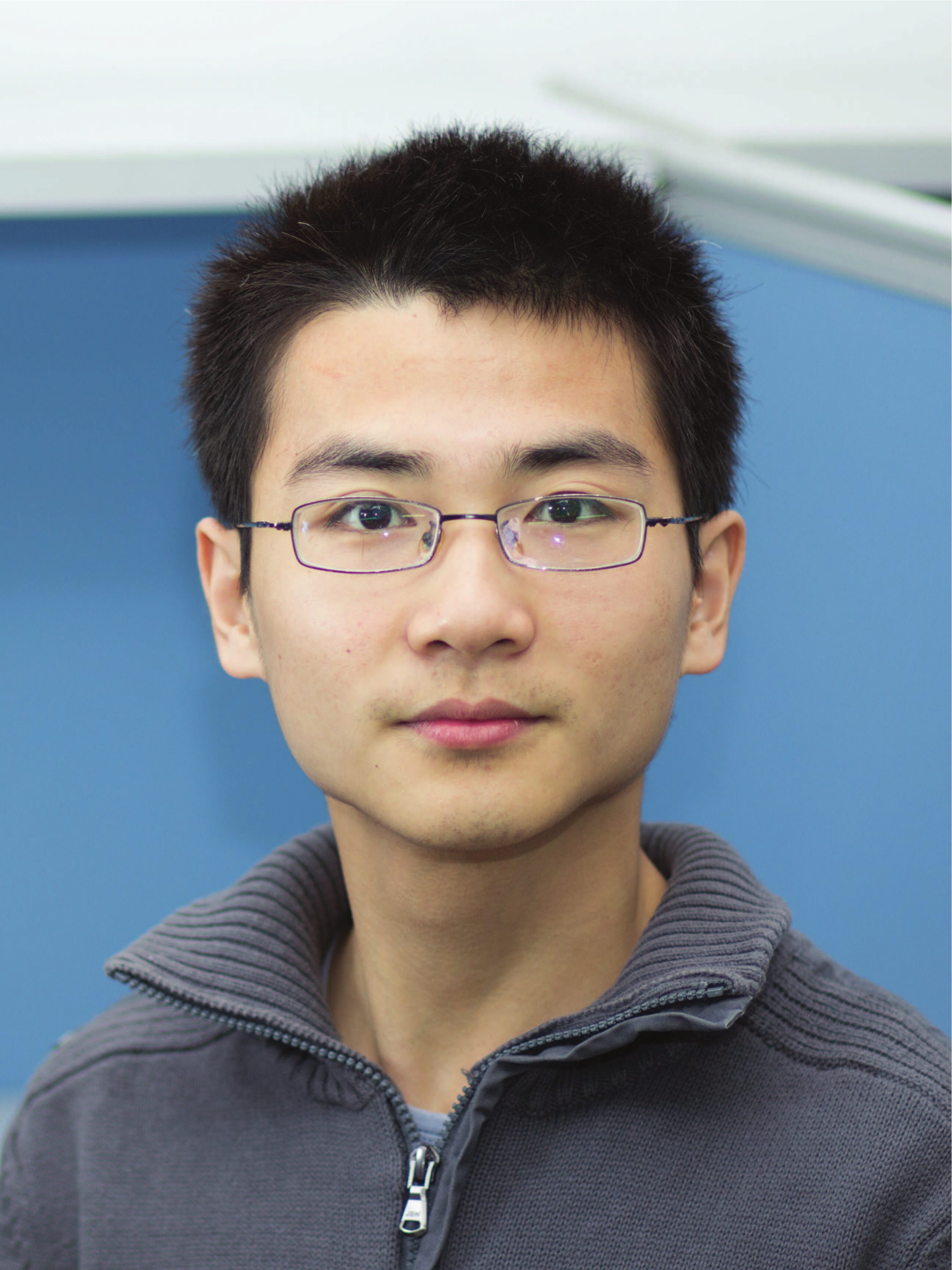}}]{Beixiong Zheng}
	(M'18) received the B.S. and Ph.D.
	degrees from the South China University of Technology, Guangzhou, China, in 2013 and 2018, respectively. 
	He is currently a Research Fellow with the Department of Electrical and Computer Engineering, National University of Singapore.
	He is also serving as an Editor for the IEEE Communications Letters.
	His recent research interests include intelligent reflecting surface (IRS), index modulation (IM), and non-orthogonal multiple access (NOMA). 
	
	From 2015 to 2016, he was a Visiting Student Research Collaborator with Columbia University, New York, NY, USA. He was a recipient of
	the Best Paper Award from the IEEE International Conference on Computing, Networking and Communications in 2016, 
	the Best Ph.D. Thesis Award from China Education Society of Electronics in 2018,
	and the Outstanding Reviewer of Physical Communication in 2019. 
\end{IEEEbiography}

\begin{IEEEbiography}[{\includegraphics[width=1in,height=1.25in,clip,keepaspectratio]{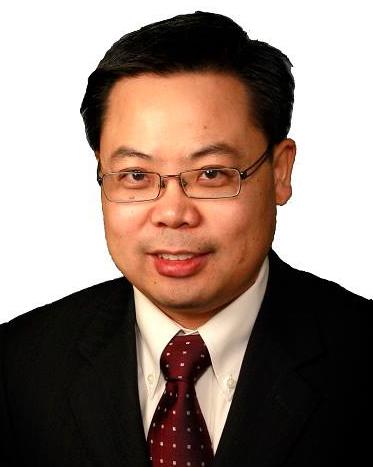}}]{Xianbin Wang (S'98-M'99-SM'06-F'17)} is a Professor and Tier-I Canada Research Chair at Western University, Canada. He received his Ph.D. degree in electrical and computer engineering from National University of Singapore in 2001.
	
Prior to joining Western, he was with Communications Research Centre Canada (CRC) as a Research Scientist/Senior Research Scientist between July 2002 and Dec. 2007. From Jan. 2001 to July 2002, he was a system designer at STMicroelectronics, where he was responsible for the system design of DSL and Gigabit Ethernet chipsets.  His current research interests include 5G technologies, Internet-of-Things, communications security, machine learning and locationing technologies. Dr. Wang has over 300 peer-reviewed journal and conference papers, in addition to 26 granted and pending patents and several standard contributions.
	
Dr. Wang is a Fellow of Canadian Academy of Engineering, a Fellow of IEEE and an IEEE Distinguished Lecturer. He has received many awards and recognitions, including Canada Research Chair, CRC President’s Excellence Award, Canadian Federal Government Public Service Award, Ontario Early Researcher Award and five IEEE Best Paper Awards. He currently serves as an Editor/Associate Editor for IEEE Transactions on Communications, IEEE Transactions on Broadcasting, and IEEE Transactions on Vehicular Technology and He was also an Associate Editor for IEEE Transactions on Wireless Communications between 2007 and 2011, and IEEE Wireless Communications Letters between 2011 and 2016. Dr. Wang was involved in many IEEE conferences including GLOBECOM, ICC, VTC, PIMRC, WCNC and CWIT, in different roles such as symposium chair, tutorial instructor, track chair, session chair and TPC co-chair.
\end{IEEEbiography}
	
\end{document}